\documentclass[aps,letterpaper,twocolumn,prl,footinbib,floatfix]{revtex4}
\usepackage{amsmath}
\usepackage{amsfonts}
\usepackage{amssymb}
\usepackage[scanall]{psfrag}
\usepackage{psfrag}
\usepackage{graphicx}
\usepackage{color}

\begin{document}
\newcommand{\of}[1]{\left( #1 \right)}
\newcommand{\sqof}[1]{\left[ #1 \right]}
\newcommand{\abs}[1]{\left| #1 \right|}
\newcommand{\avg}[1]{\left< #1 \right>}
\newcommand{\cuof}[1]{\left \{ #1 \right \} }
\newcommand{\bra}[1]{\left < #1 \right | }
\newcommand{\ket}[1]{\left | #1 \right > }
\newcommand{\pil}{\frac{\pi}{L}}
\newcommand{\bx}{\mathbf{x}}
\newcommand{\by}{\mathbf{y}}
\newcommand{\bk}{\mathbf{k}}
\newcommand{\bp}{\mathbf{p}}
\newcommand{\bl}{\mathbf{l}}
\newcommand{\bq}{\mathbf{q}}
\newcommand{\bs}{\mathbf{s}}
\newcommand{\psibar}{\overline{\psi}}
\newcommand{\svec}{\overrightarrow{\sigma}}
\newcommand{\dvec}{\overrightarrow{\partial}}
\newcommand{\bA}{\mathbf{A}}
\newcommand{\bdelta}{\mathbf{\delta}}
\newcommand{\bK}{\mathbf{K}}
\newcommand{\bQ}{\mathbf{Q}}
\newcommand{\bG}{\mathbf{G}}
\newcommand{\bw}{\mathbf{w}}
\newcommand{\bL}{\mathbf{L}}
\newcommand{\ohat}{\widehat{O}}
\newcommand{\up}{\uparrow}
\newcommand{\down}{\downarrow}
\newcommand{\MM}{\mathcal{M}}
\newcommand{\tX}{\tilde{X}}
\newcommand{\tY}{\tilde{Y}}
\newcommand{\tZ}{\tilde{Z}}
\author{Eliot Kapit}
\affiliation{Department of Physics and Engineering Physics, Tulane University, New Orleans, LA 70118}

\title{The upside of noise: engineered dissipation as a resource in superconducting circuits}

\begin{abstract}

Historically, noise in superconducting circuits has been considered an obstacle to be removed. A large fraction of the research effort in designing superconducting circuits has focused on noise reduction, with great success, as coherence times have increased by four orders of magnitude in the past two decades. However, noise and dissipation can never be fully eliminated, and further, a rapidly growing body of theoretical and experimental work has shown that carefully tuned noise, in the form of engineered dissipation, can be a profoundly useful tool in designing and operating quantum circuits. In this article, I review important applications of engineered dissipation, including state generation, state stabilization, and autonomous quantum error correction, where engineered dissipation can mitigate the effect of intrinsic noise, reducing logical error rates in quantum information processing. Further, I provide a pedagogical review of the basic noise processes in superconducting qubits (photon loss and phase noise), and argue that any dissipative mechanism which can correct photon loss errors is very likely to automatically suppress dephasing. I also discuss applications for quantum simulation, and possible future research directions.

\end{abstract}

\maketitle

\section{Introduction}

Superconducting circuits are one of the most promising platforms for quantum computing and quantum simulation \cite{devoretschoelkopf2013}. Like all candidate systems, they suffer from random noise, and through extraordinary efforts over the past two decades coherence times have increased by four orders of magnitude since the first charge qubits \cite{bouchiatvion1998,nakamurapashkin1999}. Reducing this noise below key thresholds would enable the use of digital quantum error correction \cite{fowlersurface,terhal2015} and pave the way for large scale quantum computing. Recent designs are already at or below this threshold \cite{barendskelly2014}, though further improvements are still necessary to build a fault tolerant quantum computer. And while the quest to reduce noise continues, a rapidly growing body of theoretical and experimental work has shown that engineered quantum noise, carefully tuned to complement the properties of the system, can overcome intrinsic noise sources and thus be an extremely useful tool in designing and operating quantum circuits.

Engineered dissipation can have many advantages over traditional methods of state preparation and error correction. In ordinary superconducting circuits, the system is controlled by complex digital hardware, which either runs a pre-determined set of commands or measures the circuit, processes the results algorithmically in real time, and then performs further operations accordingly. This presents substantial overhead, and the measurement process itself is often much slower than the desired manipulations of the quantum state (for modern qubits, typical measurement times are $0.5-1$ $\mu$s, whereas gates typically take $25-150$ ns). Further, in the case of measurement based feedback, the ``decisions" that the control circuit must make are very simple, at least in the small circuits currently studied. A dissipative implementation of a given feedback process, where no active measurement is required, could thus be simpler, much faster, and yield higher fidelity results (for a recent experiment comparing the two approaches, see \cite{liushankar2016}). And in the case of many-body quantum states in ``analog" quantum simulators, dissipative schemes can generate and stabilize states even when no efficient digital protocol is known. The examples in this review  demonstrate important advances in using dissipation to prepare and manipulate quantum states in superconducting devices, and cover topics ranging from simple single qubit protocols to quantum error correction and many-body physics. Taken together, they show the promise of engineered noise as a tool both for near-term experiments and beyond.

This review is divided into three parts. First, I provide an overview of the phenomenological noise model for superconducting qubits. As many of the engineered dissipation schemes discussed in this work are designed to overcome this intrinsic noise, and are carefully tailored to its important features, understanding it is crucial for what follows. While superconducting qubit implementations differ in important details the noise they suffer from is nearly always dominated by a combination of white noise photon loss and low-frequency phase fluctuations \cite{martinisnam2003,ithiercollin2005,yoshiharaharrabi2006,bylandergustavsson2011,antonmuller2012,yangustavsson2013,paladinogalperin2014,omalleykelly2015,yangustavsson2015,kellybarends2015}. As we shall see, this particular spectral composition lends itself to very efficient correction protocols that would fail in more general cases.

Second, I discuss methods for state stabilization and generation, beginning with simple applications such as driven reset protocols and Bell state generation before moving on to many-body physics and quantum simulation. I outline in detail general quantum ratcheting mechanisms that can be used to stabilize complex many-body states, and describe an efficient approximation scheme that reduces the computational complexity of simulating them. 

The third part of this article is focused on one of the most important applications of engineered dissipation, quantum error correction, and I review small logical qubit constructions as well as more exotic proposals for large scale passive error correction. I cover three paradigmatic examples of small logical qubit circuits: a passive implementation of the three qubit bit flip code \cite{kerckhoffnurdin2010,cohenmirrahimi2014,kapithafezi2014}, cat codes \cite{leghtaskirchmair2013,mirrahimileghtas2014,sunpetrenko2014,leghtastouzard2015,albertshu2016,ofekpetrenko2016,cohensmith2016,mundhadagrimm2016,michaelsilveri2016,wanggao2016,heeresreinhold2016,puriblais2016} and the Very Small Logical Qubit (VSLQ) \cite{kapit2016,kapitetg2017}. Cat codes and the VSLQ are particularly exciting in the near term; cat codes have already demonstrated the first QEC code to exceed break-even \cite{ofekpetrenko2016}, and with minimal device counts of two resonators and a qubit, or two qubits and two resonators, respectively, they are far smaller than the smallest surface code implementation \cite{fowlersurface} of 17 qubits and a similar number of resonators. I also describe more complex models, where simulating topological states can be used to perform large scale autonomous error correction of long chains of errors. Following this discussion, I offer concluding remarks and suggest future directions for the field.

\section{Noise model for superconducting circuits}

Before delving into the myriad uses of engineered noise in superconducting qubits, we first discuss basic features of noise in superconducting qubits. While a number of competing qubit designs exist \cite{devoretschoelkopf2013}, the most successful (defined here as having high coherence and gate fidelities; the strongly coupled, low coherence flux qubits used in quantum annealing \cite{albashlidar2017} serve different purposes and are not the subject of this review) share a broad set of common features. The bare excitation energy $\omega_{01}$ of the qubit is typically a few GHz, and thus much larger than the ambient temperature of the environment (15-25 mK). The excitation energy $\omega_{12}$ to access the second excited state is detuned from $\omega_{01}$ by a substantial nonlinearity of at least a hundred MHz. These qubits are coupled to linear resonators, with similar frequency ranges.
The primary noise channels for these qubits are photon losses and phase fluctuations \cite{martinisnam2003,ithiercollin2005,yoshiharaharrabi2006,bylandergustavsson2011,antonmuller2012,yangustavsson2013,paladinogalperin2014,omalleykelly2015,yangustavsson2015,kellybarends2015}.

Photon loss, where a photon spontaneously leaks into the environment, relaxing the qubit from $\ket{1}$ to $\ket{0}$, has a noise spectrum that is roughly energy independent (see for example the data in \cite{kellybarends2015}), aside from spurious resonances with environmental degrees of freedom. To good approximation this noise is Markovian and can be modeled through non-unitary Lindblad collapse operators \cite{gardinerzoller,plenioknight1998,daley2014} $\sqrt{\Gamma_P} a$, where $\Gamma_P = 1/T_{1}$ is the photon loss rate. Since photon loss has a white noise spectrum it cannot be easily suppressed (in contrast to the phase noise discussed below), as driving schemes that effectively sample the noise spectrum at higher frequencies \cite{martinisnam2003,yangustavsson2015} do no good if the loss rate does not decrease at higher frequencies. Another way of viewing this situation is to note that while the average qubit energy is a few GHz, qubit-qubit coupling energies are at most a hundred MHz, meaning that multi-qubit effects will only shift the photon energy by a few percent. If we assume that the loss occurs from weak interactions with a broad bath of environmental degrees of freedom, it is unlikely that shifting the energy by a few percent should significantly change the environmental density of states, leading us to conclude that the photon loss rate will be roughly independent of many-body effects. Note of course that narrow, spurious resonances with environmental degrees of freedom can enhance the loss rate dramatically near particular frequencies. Away from these points, typical photon loss rates for well-designed qubits are in the $10-100$ kHz range.

The primary contributions to the photon loss rate are interactions with the broader environment and, in many cases, the off-resonant coupling to the readout resonator used to measure the qubit state; as the readout is strongly coupled to the external environment even weak coupling between it and the qubit can lead to significant noise. Photon loss through the readout can be suppressed through clever filter design \cite{reedjohnson2010,bronnliu2015}, or redesigning the readout scheme so that the static coupling between the qubit and readout is negligible \cite{didierbourassa2015}. Qubits with multiple degrees of freedom \cite{srinivasanhoffman2011} may have enhanced loss rates in some excited states but not others, due to primarily to symmetries changing how each state interacts with its environment. Due to increased environmental isolation qubits embedded in 3d cavities \cite{rigettigambetta2012,jinkamal2015} typically have lower loss rates and experience weaker phase noise, though they are more difficult to manipulate and the large 3d cavities present significant packaging challenges in multi-qubit circuits. 

Resonators also experience photon loss; planar resonators not coupled to a readout typically have loss rates similar to those of transmon qubits, though 3d cavity resonators can have much lower loss rates, with $T_1$ in the 1 to 10 ms range achieved in experiments \cite{reagorpaik2013,reagorpfaff2016}. Finally, photon addition is rare due the very low environmental temperature (relative to the qubit energy) \cite{jinkamal2015} and is very weak compared to losses, at least at current coherence levels \footnote{This is not to say that photon addition is in all cases ignorable. For reasons that are somewhat unclear, in many experiments the ambient photon population is much higher than the environmental temperature would suggest, though still small enough that the incoherent photon addition rate is orders of magnitude smaller than the loss rate. These ``thermal" photons can create errors either through directly exciting the qubit, or its readout resonator, which causes phase noise through the dispersive coupling and reduces $T_{2}$. While these effects are largely irrelevant in most of the systems considered in this review, they are present and could become a limiting noise source in the future.}. In summary, the basic picture to keep in mind for photon loss is white noise, non-unitary amplitude damping, which cannot be suppressed by multi-qubit effects but can be repaired through error correction schemes.

In contrast to the white noise spectrum of losses, phase noise is low-frequency dominated \cite{martinisnam2003,ithiercollin2005,yoshiharaharrabi2006,bylandergustavsson2011,antonmuller2012,yangustavsson2013,paladinogalperin2014,omalleykelly2015,yangustavsson2015}, with a spectrum composed of $1/f^{\alpha}$ (where $\alpha \simeq 1$; for simplicity we will refer to this as $1/f$ noise from now on) and telegraph \cite{paladinogalperin2014} components. Both of these spectra vanish in the high frequency limit and are strongest as $\omega \to 0$, and can thus be suppressed by continuous driving and spin echo procedures, which sample the noise at higher frequencies and thus diminish its effect on a quantum state. Weak but nonvanishing high frequency phase noise has been observed in flux qubit experiments \cite{yangustavsson2013,yangustavsson2015}, though not in transmons. Lacking flux tunability and critical current noise, resonators typically experience much weaker phase noise than qubits.

White noise amplitude damping and low-frequency phase noise thus comprise the basic noise model of superconducting qubits, and overcoming them is the central challenge in building a quantum computer. The engineered dissipation schemes discussed in this review present a number of ways to do so, and have applications in both quantum simulation and digital quantum computing.

\section{State generation and stabilization}

One of the most important and obvious uses of engineered dissipation is to prepare and stabilize quantum states. State preparation is fundamental to both quantum computing and quantum simulation, and engineered dissipation is a particularly simple and efficient way to prepare states, ranging from simple single qubit configurations to dissipative entanglement generation and even true many-body state stabilization. We note also that dissipation can be used more generally to implement gates \cite{verstraete,kastoryanowolf2013} and improved qubit measurement protocols \cite{didierbourassa2015,voolshankar2016}, among other things. For the moment, however, we will focus on stabilization.

Perhaps the simplest form of stabilization is cooling, both in sense of the physical removal of energy of the system and in relaxing the system toward a dark state in continuously driven multi-qubit setups. And beyond the obvious desire one might have to reduce a system's temperature for various purposes, engineered physical cooling (in the rest frame) is particularly useful in quantum information processing. Quantum error correction codes \cite{fowlersurface,terhal2015}, generally thought to be the most promising route toward a truly fault-tolerant quantum computer, require large numbers of ancilla qubits to make repeated multi-qubit measurements, and in order to avoid continuous error propagation scrambling the measurement sequence, these qubits must be reset to a known state (typically the rest frame ground state $\ket{0}$) at the beginning of each EC cycle. Consequently, methods to rapidly reset qubits can shorten the EC cycle, and engineered dissipation is a natural and hardware-efficient way to do so.

\begin{figure}
\includegraphics[width=3.5in]{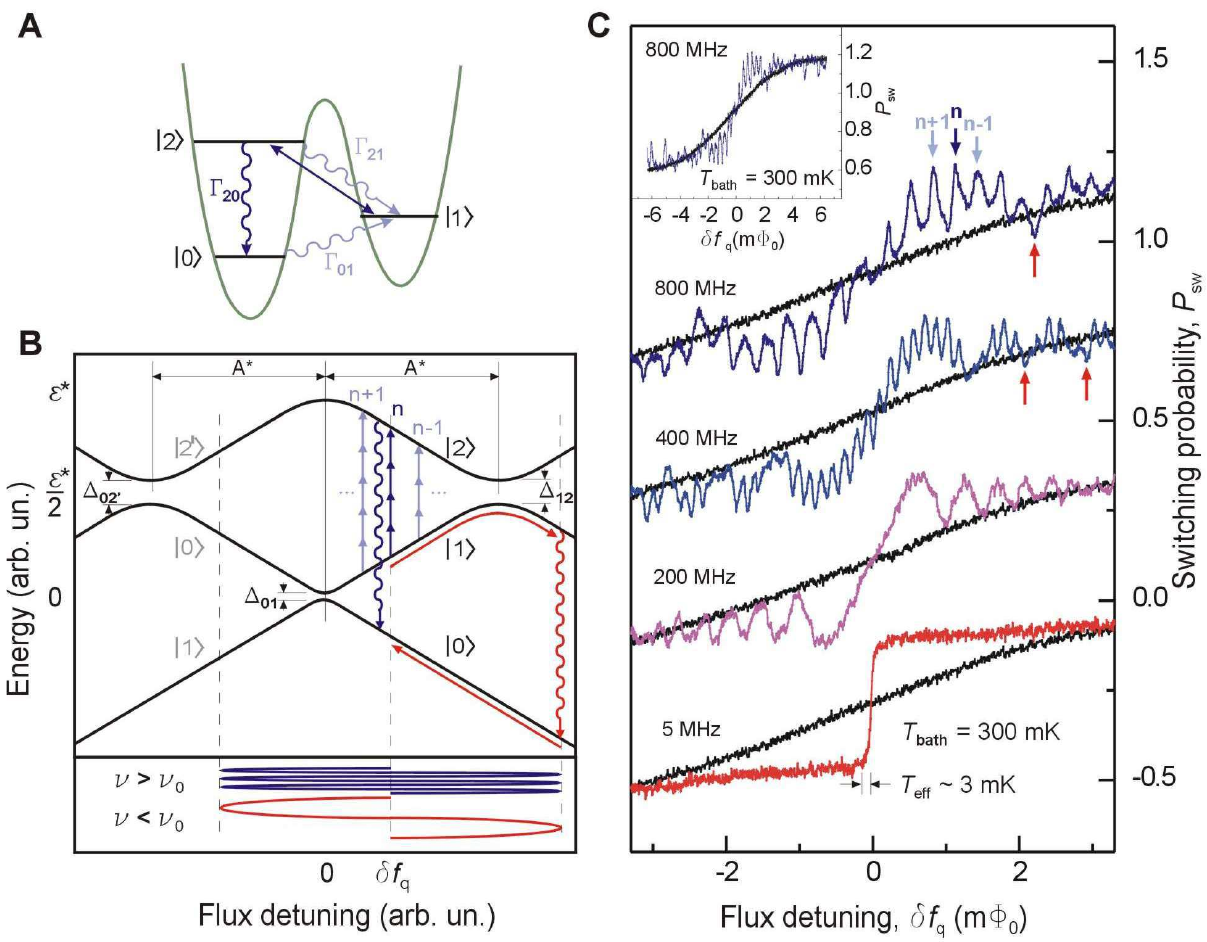}
\caption{Microwave cooling of a superconducting flux qubit, reproduced with permission from \cite{valenzuelaoliver2006}. The flux qubit is operated as a three-level system (parts A and B), where states $\ket{1}$ and $\ket{2}$ can both decay to $\ket{0}$, but the rate for $\ket{2} \to \ket{0}$ is much faster than the rate for $\ket{1} \to \ket{0}$. Consequently, by resonantly driving the flux qubit's $\ket{1} \to \ket{2}$ transition, the effective decay rate from $\ket{1} \to \ket{0}$ can be dramatically increased. This in turn reduces the effective temperature of the flux qubit (set by the average population of the $\ket{1}$ state, part C) to levels well below the ambient temperature of the system. Driven qubit reset protocols such as this one may be important for engineering a fault-tolerant quantum computer.}\label{valenzuelafig}
\end{figure}

One of the first important demonstrations of engineered dissipation in superconducting circuits was the work of Valenzuela \textit{et al} \cite{valenzuelaoliver2006}, who cooled a superconducting flux qubit well below the ambient temperature of its environment using continuous driving, shown in FIG.~\ref{valenzuelafig}. In this scheme, the flux qubit has a biased double-well potential with three relevant energy levels. These levels are the true ground state $\ket{0}$, the first excited state $\ket{1}$ (which corresponds to occupation of the opposite well, and thus decays weakly due to the large potential barrier suppressing inter-well tunneling) and the second excited state $\ket{2}$, which is an excitation in the same well as $\ket{0}$. There is a large nonlinearity such that $\of{\omega_2 - \omega_1}$ and $\of{\omega_1 - \omega_0 }$ are well-separated. In the absence of any driving the system remains mostly in $\ket{0}$ with a small probability $P_1 \propto e^{- \hbar \omega_1 / k_{B} T}$ of being excited by exchange of thermal photons with the environment. This probability can be reduced still further by applying an AC flux tone at $\omega_{2} - \omega_1$, such that whenever the qubit is excited it resonantly mixes with $\ket{2}$, which then decays rapidly back to $\ket{0}$. When applied continuously this reduces the excitation probability $P_1$ -- and thus, the effective temperature of the qubit-- dramatically, with effective temperatures below 3 mK achieved for very strong driving even when the ambient environment was held at 300 mK. As in many other examples in this review, when the driving is strong enough heating from off-resonant transitions limits the minimum achievable temperature. Though simple in nature this pioneering experiment was one of the first demonstrations of the power of engineered dissipation for state stabilization in superconducting devices.

Similar protocols are now used widely for state reset in other experiments. A more recent highlight is the DDROP protocol introduced by Geerlings et al \cite{geerlingsleghtas2013}, which allows for the rapid reset of a transmon qubit dispersively coupled to a lossy readout resonator, and since capacitive coupling to a detuned resonator is the generic method for measuring the state of a transmon, it can be applied widely. A parallel scheme by McClure \textit{et al} \cite{mcclurepaik2016} can be used to rapidly reset a readout resonator, achieving a decay rate twice that of natural decay through pulse shaped driving. Minimizing stray photons in readout resonators suppresses a possible dephasing channel, and while photons remain in the resonator they continuously interact with the qubit, preventing its use in further gates until the resonator empties. Methods for rapidly resetting qubits and resonators are thus invaluable for engineering digital error correction codes.

These examples focus on enhancing natural relaxation processes to reach the ground state of a system more rapidly and with higher fidelity. Engineered dissipation can also be used to stabilize the excited states of qubits and resonators against photon losses. The first demonstration of state stabilization in a superconducting qubit through driving and dissipation was by Murch \textit{et al} \cite{murchvool2012}, who used continuous driving and dissipation to relax a qubit toward chosen superpositions of ground and excited states. In this method a qubit is driven on resonance and dispersively coupled to a detuned cavity, producing a system Hamiltonian $H = - \Omega_R \sigma^x /2 - \chi a^\dagger a \sigma^z$. A second drive is added to the system that drives the cavity detuned by an amount $\Delta_c$. Given a cavity damping rate $\kappa$ and average population $\bar{n}$, as a function of $\Delta_c$, the net heating ($\Gamma_+$) and cooling ($\Gamma_-$) rates in the qubits rotating frame can be calculated to be
\begin{eqnarray}
\Gamma_\pm = \frac{\kappa \chi^2 \bar{n}}{ \of{\Omega_R \pm \Delta_c}^2 + \kappa^2 /4 } + \frac{1}{2 T_2 }
\end{eqnarray}
Here, $T_2 = \of{ 1 / 2 T_1 + 1/ T_\varphi}^{-1}$ is the qubit's phase coherence time. Inspection of the rates $\Gamma_{\pm}$ shows that if we pick the detuning $\Delta_c = \pm \Omega_R$ one rate will be maximal while the other will be small, causing the system to relax toward one of the two rotating frame eigenstates. The experimenters observed a maximum fidelity of approximately 0.7, primarily limited by readout infidelity and mixing with the $\ket{2}$ state of the transmon. Beyond the novelty of passively stabilizing a coherent superposition state, the basic mechanism-- correcting errors by matching the detuning of a driven lossy element to the Rabi frequency of a drive term in the primary system-- is fundamental to the passive error correction schemes detailed later in this review. Further refinements of this setup allow for the simultaneous measurement of non-commuting observables \cite{hacohenmartin2016}.

More recently, Holland \textit{et al} used dissipative techniques to stabilize a single photon Fock state in a superconducting cavity \cite{hollandvlastakis2015}. Their scheme used a pair of superconducting cavities (one storage, one lossy for cooling/readout) and a transmon qubit, which induced a cross-Kerr nonlinear interaction between the cavities, and is methodologically similar to that of Geerlings \textit{et al}. In their experiment they achieved a one-photon steady state probability of 0.63, corresponding to an effective negative temperature of -0.77K. This experiment was also the first to demonstrate strong enough cross-Kerr interactions to resolve single photon states, providing a valuable tool for engineering novel cavity dynamics. 

\subsection{Entanglement generation}

\begin{figure}
\includegraphics[width=3.5in]{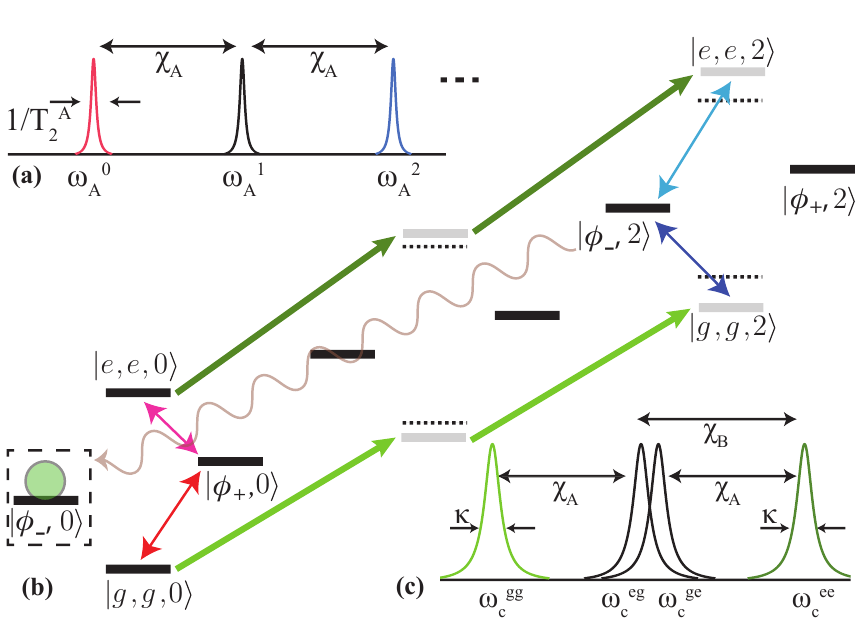}
\caption{Autonomously stabilizing a Bell state in a pair of qubits sharing a common cavity, reproduced with permission from \cite{leghtasvool2013}. As discussed in the text, six drive tones are applied to a pair of qubits dispersively coupled to a cavity, with tuned frequencies and phases such that the antisymmetric Bell state $\ket{\phi_-} = \of{\ket{01} - \ket{10}}/\sqrt{2}$ is detuned from all drives and unaffected by them, but all other states are rapidly mixed with configurations that contain $\ket{\phi_-}$ as a decay outcome. The system thus quickly decays to $\ket{\phi_-}$ from any initial condition, with a steady state fidelity of up to 67\% reported in the experiment (rising to 77\% with postselection).}\label{leghtas2013fig}
\end{figure}

Having discussed driven cooling and single qubit state stabilization, we now move on to the more complex topic of entanglement generation, where engineered dissipation is used to stabilize entangled states of multiple degrees of freedom \cite{leghtasvool2013,shankarhatridge2013,aronkulkarni2014,schwartzmartin2016,liushankar2016}. The simplest and most experimentally studied case is the stabilization of a Bell state of two qubits, $\ket{\phi_{\pm}} = \of{\ket{01} \pm \ket{10} }/\sqrt{2}$.

Following \cite{leghtasvool2013,shankarhatridge2013}, we consider a pair of qubits $A$ and $B$ capacitively coupled to a common cavity, as shown in FIG. \ref{leghtas2013fig}. The qubit frequencies $\omega_A$ and $\omega_B$ are well detuned from the cavity frequency $\omega_c$ and from each other, so the net effect is to engineer a dispersive interaction Hamiltonian $H = \of{\chi_A a_A^\dagger a_A + \chi_B a_B^\dagger a_B } a_c^\dagger a_c$. The energies and couplings are chosen so that $\chi_A \simeq \chi_B = \chi$; it is this nonlinear interaction that allows the entangled state to be stabilized. To stabilize the bell state $\ket{\phi_-} = \of{\ket{01} - \ket{10} }/\sqrt{2}$, we apply a total of six drive tones. The first pair, at frequencies $\omega_c^{gg}$ and $\omega_c^{ee}$, are targeted at the resonator energy when the qubits are in $\ket{00}$ or $\ket{11}$, respectively; these drive the cavity to a coherent state if the qubits are in $\ket{00}$ or $\ket{11}$ but are off resonant and have negligible effect when the qubits are in any combination of $\ket{01}$ and $\ket{10}$. We next apply a pair of drives at frequencies $\omega_A$ and $\omega_B$ with equal amplitudes $\Omega_0$, which mixes $\ket{\phi_+}$ with $\of{\ket{00} + \ket{11}}/\sqrt{2}$ but leaves the $\ket{\phi_-}$ states alone. Finally, to isolate $\ket{\phi_-}$, we apply two more drives at frequencies $\omega_A + \bar{n} \chi$ and $\omega_B + \bar{n} \chi$, with the same Rabi frequency but locked phases to isolated the antisymmetric state. These drives will resonantly mix $\ket{00,n}$ and $\ket{11,n}$ with $\ket{\phi_-, n}$, where $n$ is the closest integer to the average photon number in the cavity. However, once the system is in $\ket{\phi_-,n}$ it rapidly decays to $\ket{\phi_-,0}$, and due to the resonance condition this state is unaffected by \textit{any} of the six drives. In contrast to all the other states, it is stable aside from losses due to finite qubit $T_1$, and thus becomes the steady state of the system, with a fidelity of up to 67\% achieved in the experiment.

It is important to notice that the only irreversible step in this protocol is the cavity decay from $\ket{\phi_-, n}$ to $\ket{\phi_-,0}$. As in the flux qubit cooling described earlier \cite{valenzuelaoliver2006}, $\ket{\phi_-,0}$ is stabilized energetically: once in this state the system is off-resonant with all drives, but when a decay occurs the system enters a state which is rapidly pumped into an unstable manifold that has $\ket{\phi_-,0}$ as one of its decay outcomes, and since this is the only long-lived state the system will rapidly relax toward it from all other configurations. This process is a type of ``quantum ratcheting" \cite{reimanngrifoni1997}, with resonance conditions controlling the relative stability of the target state; once decay processes populate the target state it is off resonant with all drives that would mix it with other states, ensuring its stability. We will present a more general picture of cooling based on resonance conditions shortly.

Demonstrating the power of engineered dissipation, the Yale team also compared this method of driven-dissipative entanglement stabilization to an active, measurement-based scheme \cite{liushankar2016}. Using a new sample, they compared the previously described driven-dissipative protocol to a protocol where an FPGA-controlled circuit performs repeated ``quasiparity" measurements to distinguish 0 and 2 photon states from the single photon manifold; if a 0 or 2 photon state is detected a pulse is applied which mixes $\ket{00}$ and $\ket{11}$ with $\ket{\phi_-}$ with 50\% probability. If on the other hand a single photon state is detected a pulse is applied which leaves $\ket{\phi_{-}}$ invariant but mixes $\ket{\phi_+}$ with $\ket{00}$ and $\ket{11}$, which is then detected in the next measurement cycle and has a 50\% chance of being corrected to $\ket{\phi_-}$. The steady state fidelities of $\ket{\phi_-}$ are 76\% for driven-dissipative methods and 58\% for the measurement scheme. The measurement scheme has worse performance primarily due to its comparatively long cycle time ($1.5 {\rm \mu s}$); in the driven dissipative method the effective decay time $\tau$ toward the $\ket{\phi_-}$ state is only $0.78 {\rm \mu s}$, or about twice as fast. We will see later that the comparatively high speed of dissipative processes can be used for passive quantum error correction that outperforms measurement-based codes for small systems. The authors further improved the performance by turning on and off the dissipative drives based on a set of additional measurements; the steady state fidelity improved to 82\% in that case. These works elegantly demonstrate how a combination of symmetries and energetic resonance conditions can stabilize entangled quantum states.

Other approaches to dissipatively stabilizing entanglement have been recently reported. For example, as proposed in \cite{aronkulkarni2014}, and realized experimentally by \cite{schwartzmartin2016}, two qubits are prepared with near-identical resonance energies and coupled to each other through an effective exchange coupling $J \of{\sigma_A^+ \sigma_B^- + \sigma_A^- \sigma_B^+}$, mediated by coupled cavities. This creates a splitting between $\ket{\phi_+}$ and $\ket{\phi_-}$, and by carefully tuning drives and dissipation to narrow resonance with $\ket{\phi_-}$ or $\ket{\phi_+}$ the two states can be distinguished, with an experimentally reported steady state fidelity of 70\%. This protocol is notable because the cavities containing each qubit play a triple role in inducing an interaction between qubits, serving as the lossy element for quantum ratcheting, and acting as readout elements. Further, they can be spatially separated by potentially large distances, so long as the coupling between them is appreciable. In both this protocol and the one studied by the Yale group dephasing reduces the fidelity by mixing $\ket{\phi_\pm}$, though it can be suppressed through additional drive tones \cite{heinaron2016}. Modest improvements in qubit coherence times and refinements to the protocols of this section could achieve fidelities better than 90\%, creating an on-chip source of entangled states for more complex circuits.

\subsection{Stabilizing many-body states}

Following the dictum ``one, two, many," we now turn to many-body physics. While predicting the dynamics of a generic model interacting with a colored bath is an extremely difficult problem as argued below, if the many-body system under consideration is well-understood one can often tailor a bath to generate and stabilize its eigenstates. Such bath engineering has a rich history in AMO systems (see for example \cite{poyatoscirac1996,diehlmicheli2008,verstraete,barreiromuller2010,krautermuschik2011,lingaebler2013,budichzoller2015,morigieschner2015,kaczmarczykweimer2016}), 
and it is extremely promising in superconducting circuits as well, given that couplings, energies and dissipation rates can all be tuned over wide ranges in sub-$\mu$s timescales. In particular, if the eigenvalues and eigenvectors of the primary system are well known \footnote{Importantly, one does not need to know the entire spectrum of an interacting many-body system to stabilize its ground state. Instead, one must simply have a good understanding of all states that can be reached in a few-error process from the target state, and of all states on or near (in number of events) the most likely dissipative path between a chosen starting state and the target state. With this information one can design a bath that drives the system from the starting configuration to its target state, though complications such as energetic frustration and local energetic minima may make this very difficult in practice.} one can often design a bath that will relax the system toward the many-body interacting ``ground" state (of the driven, rotating-frame Hamiltonian). Good candidates for passive stabilization are generally gapped, strongly interacting systems; these include exotic topological models, which we will discuss later in the section concerning passive quantum error correction. The stabilization schemes in this section all focus on photon loss, and not dephasing, though that too can be corrected passively through similar methods. Further, many of these models will passively suppress dissipation, for reasons that will become evident in our consideration of passive quantum error correction.

Anticipated by our earlier discussion of one- and two-qubit physics, an extremely general mechanism for stabilizing general many-body states against photon loss is quantum ratcheting based on energetics. Imagine that we have a gapped quantum many-body system prepared in its rotating frame ground state, which we want to stabilize against photon losses; stabilizing nearly gapless states is much more difficult and beyond the scope of this review. In this frame photon losses create finite energy ``hole" excitations, and let us assume for simplicity that these excitations have a single excitation energy $E$. Over time, photon losses will populate the system with excitations, and resonant driving at $E$ will not help prevent this, since the steady state population of a given excited state relative to the ground state under resonant driving is 50\%, and thus at best driving will produce a ground state population that scales as $2^{-N}$, where $N$ is the number of qubits suffering losses (further, if the excitations have energetic dispersion or interact the rotating frame ground state population can be much worse than this). This is because resonant driving is a unitary process that is just as likely to take a system from $\ket{0}$ to an excited state $\ket{E}$ as it is from $\ket{E}$ to $\ket{0}$. To stabilize the ground state, we couple the system to a lossy qubit or resonator $S$, with excitation energy $\omega_S$. If we then drive a \textit{two}-photon transition that simultaneously returns $\ket{E} \to \ket{0}$ and excites $S$ (at frequency $\omega_S - E$), whenever an excitation is created it will Rabi-flop between the many-body system and the lossy object, where it rapidly relaxes, having removed the hole excitation and returned the system to its ground state. The ratcheting effect occurs if we assume that the two-photon matrix element $\Omega$ is small compared to $E$; in this limit creating further excitations is far off resonant, so the two-photon drive only acts when it is replacing a photon that has been lost. 

\begin{figure}
\includegraphics[width=3.0in]{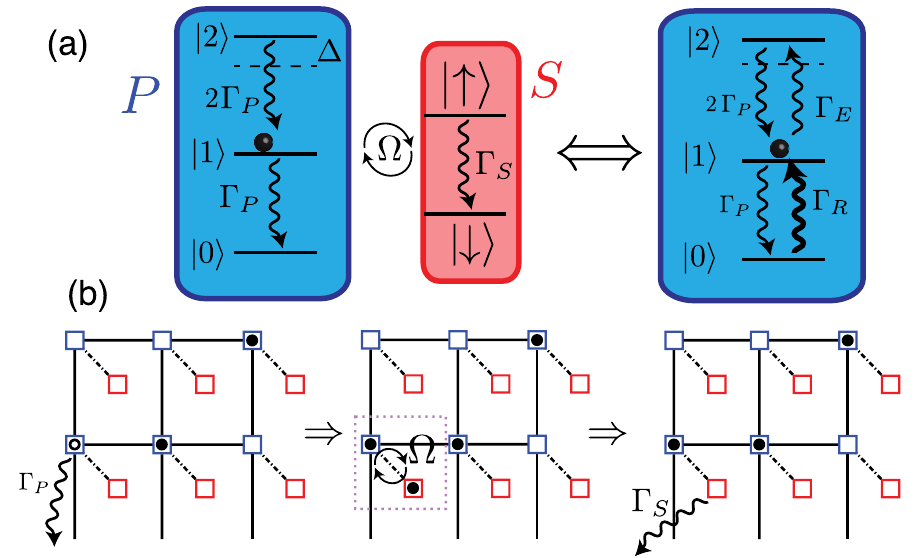}
\caption{Schematic picture of state stabilization in a simple three-level system, reproduced with permission from \cite{kapithafezi2014}. As described in the text, a parametric two-photon drive refills lost photons via coupling to an auxiliary, lossy subsystem, making the $\ket{1}$ photon state the most stable state of the system. In (b), the same process is generalized to a 2d quantum many-body system. By coupling the primary lattice (blue squares) to a lossy ``shadow" lattice (red squares), many body hole excitations created by photon losses can be passively removed.}\label{Rrate}
\end{figure}

To demonstrate this explicitly and show how the cooling or repair rates can be computed, consider a simple three-level system $P$ (e.g. a transmon qubit), as shown in FIG.~\ref{Rrate}, with loss rate $\Gamma_P$ and nonlinearity $\Delta$, coupled by a two photon driving process to a lossy two-level system $S$, with loss rate $\Gamma_S$. We want to maximize the population of the $\ket{1}$ state of $P$ (and thus minimize the population of the $\ket{0}$ and $\ket{2}$ state), and we do so by resonantly driving the two-photon coupling at $\omega_P + \omega_S - \nu$. The total Hamiltonian is
\begin{eqnarray}\label{H3}
H = \Delta \ket{2_P} \bra{2_P} + \Omega \of{a_P^\dagger a_S^\dagger + a_P a_S} + \nu a_S^\dagger a_S
\end{eqnarray}
We can now compute the rates of transition in $P$ from Fermi's golden rule, effectively integrating out the lossy $S$ degree of freedom. Noting that its loss rate $\Gamma_S$ gives it a Lorentzian density of states, we first compute the rate of inducing a transition in $P$ and exciting $S$:
\begin{eqnarray}
\gamma_{ij} =  \frac{\Omega^2 M_{ij}^2 \Gamma_S}{\of{\delta E + \nu}^2 + \Gamma_S^2 / 4  }
\end{eqnarray}
Here $M_{ij}$ is a matrix element ($M_{01} = 1$, $M_{12} = \sqrt{2}$) and $\delta E$ is the change in energy of the primary system. For the process to complete, the photon must decay (rate $\Gamma_S$), so the \textit{total} transition rate is $\Gamma_{ij} = \of{ \gamma_{ij}^{-1} + \Gamma_S^{-1} }^{-1}$. Let $\Gamma_R$ be the ``repair" rate of the transition from $\ket{0_P}$ to $\ket{1_P}$ and $\Gamma_E$ be the induced ``error" rate of the $\ket{1_P}$ to $\ket{2_P}$ transition. Explicitly:
\begin{eqnarray}\label{GRE}
\Gamma_R = \frac{\Omega^2 \Gamma_S}{\nu^2 + \Omega^2 + \Gamma_S^2 /4 }, \; \Gamma_E = \frac{2 \Omega^2 \Gamma_S}{\of{\nu+\Delta}^2 + 2 \Omega^2 + \Gamma_S^2 /4 }
\end{eqnarray}
The above expressions can be straightforwardly applied to much more complex systems by simply modifying the matrix element $M_{ij}$ and energy change $\delta E$ accordingly. Provided the nonlinearity $\Delta$ is large compared to $\Omega$ and $\Gamma_P$, and the detuning $\abs{\nu} \leq \Omega$, the system can be maintained in $\ket{1}$ to high fidelity, with $P_{1} \simeq 1 - \Gamma_P / \Gamma_R - \Gamma_E/ 2 \Gamma_P$. 

While the approximation of treating the $PS$ coupling as an incoherent transition ignores coherent quantum dynamics and thus can be inaccurate at short times, the intermediate and long time behavior is quantitatively captured by this simple calculation quite accurately in many systems. And indeed, as discussed in \cite{kapithafezi2014}, this method can be used to integrate out dissipative elements in larger many-body systems as well. Let $\ket{\bar{j}}$ be the eigenbasis of the primary system (in this case, the three-level transmon). Let us define the (dimensionless) modified creation operators $\tilde{a}_P$ by:
\begin{eqnarray}
\of{\tilde{a}_P^\dagger}_{ij} \equiv  \frac{ \Omega }{\sqrt{\of{E_i - E_j + \nu}^2 + \Gamma_S^2 / 4 + \Omega^2}  } \of{a_P^\dagger}_{ij} .
\end{eqnarray} 
Here $\nu$ is the energy of the lossy degree of freedom $S$ in the rotating frame, $E_i$ and $E_j$ are the energies of levels $i$ and $j$, and $\Omega$ is the coupling matrix element. The transmon's density matrix $\rho$ then evolves under the modified Lindblad equation
\begin{eqnarray}
\partial_t \rho &=&  \frac{i}{\hbar} \sqof{\rho,H} + \Gamma_P \of{a_P \rho a_P^\dagger + \frac{1}{2} \of{a_P^\dagger a_P \rho + \rho a_P^\dagger a_P}  } \nonumber \\
& & + \Gamma_S \of{\tilde{a}_P^\dagger \rho \tilde{a}_P + \frac{1}{2} \of{\tilde{a}_P \tilde{a}_P^\dagger \rho + \rho \tilde{a}_P \tilde{a}_P^\dagger}  }.
\end{eqnarray}
This expression is easier to evaluate numerically since the density matrix being evolved consists of only the transmon and not the transmon and lossy object. It can also be scaled up to larger systems with multiple degrees of freedom and dissipation sources-- one merely sums over the above expression for each individual incoherent term. For an $N$ qubit system with identical decay rates $\Gamma_P$ coupled to $N$ identical lossy elements coupled to it we would have:
\begin{eqnarray}
\partial_t \rho &=& \frac{i}{\hbar} \sqof{\rho,H} + \Gamma_P \sum_{n=1}^N \of{a_{nP} \rho a_{nP}^\dagger + \frac{1}{2} \cuof{a_{nP}^\dagger a_{nP}, \rho }  } \nonumber \\
& & + \Gamma_S \sum_{n=1}^N \of{\tilde{a}_{nP}^\dagger \rho \tilde{a}_{nP} + \frac{1}{2} \cuof{\tilde{a}_{nP} \tilde{a}_{nP}^\dagger, \rho }  }.
\end{eqnarray}
Further dissipation sources can be treated by additional terms in the sum on the second line. Given the substantial Hilbert space reduction, this method allows for the numerical simulation of larger systems than one would be able to study taking the full quantum dynamics of the lossy elements into account.

This scheme is simple enough to be quite generic, and if the primary Hamiltonian does not conserve total photon number (as in the topological models discussed toward the end of this review), photon losses can be corrected through more general couplings, including the simple capacitive coupling discussed in the three-qubit bit flip code. In addition, if other means exist to distinguish states (such as symmetries), one can arrange interactions to achieve a ratcheting effect even when the energy scales themselves are too weak to resolve the desired transition with sufficient speed. Quantum ratcheting will be illustrated in further detail through specific examples throughout this review. 

\begin{figure*}
\includegraphics[width=6.5in]{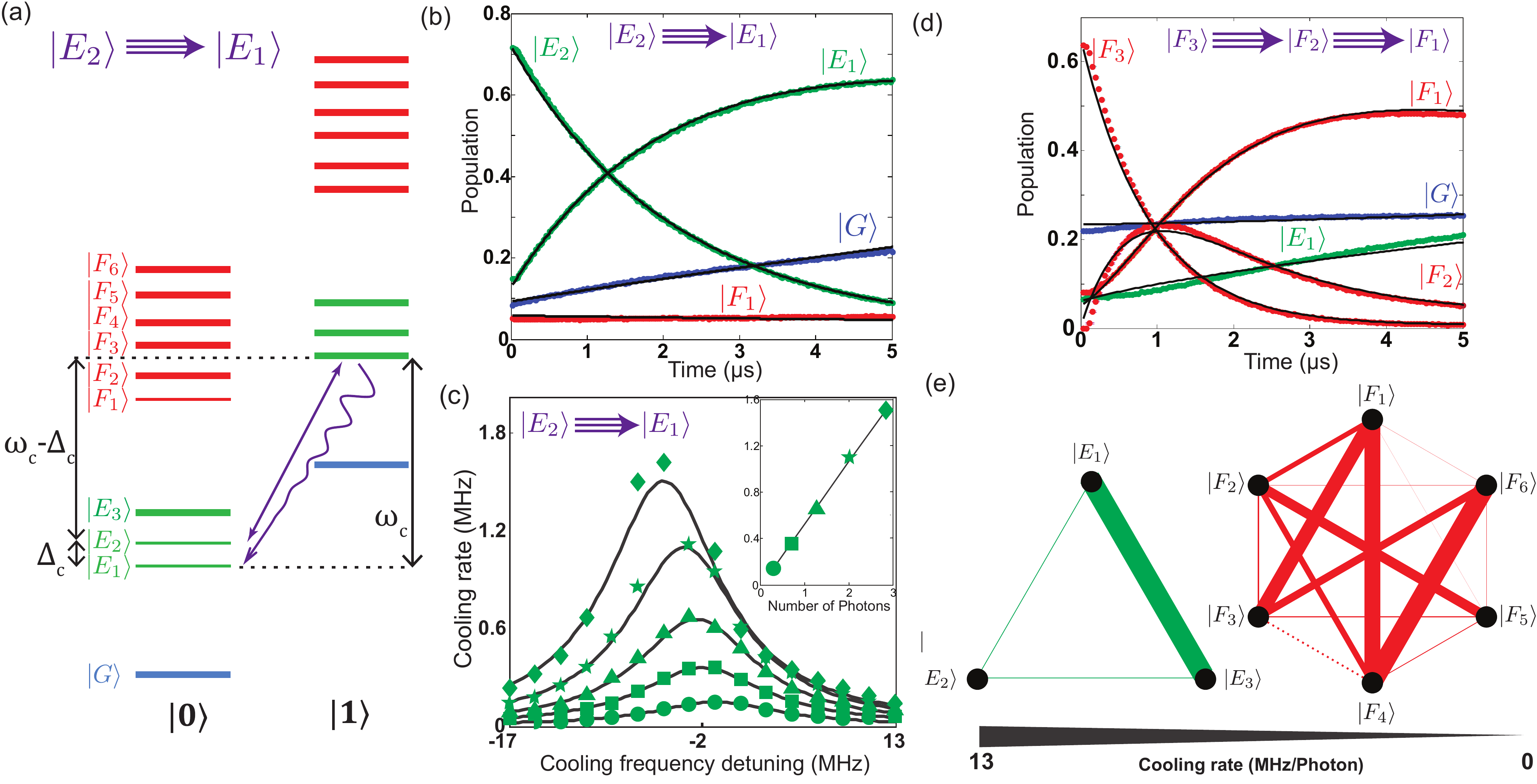}
\caption{Experimental demonstration of state stabilization through engineered loss in a 3-qubit Bose-Hubbard chain, reproduced with permission from \cite{gourgyramasesh2015}. The system is comprised of three transmon qubits and three resonators, and due to its simplicity its entire spectrum (a) can be easily computed. Armed with this knowledge, the system can be made to relax into chosen eigenstates by applying multiple drive tones (b,d). For example, in part (b) the system is driven between two single photon states $\ket{E_2}$ and $\ket{E_1}$ by resonantly driving transition between $\ket{0}$ and $\ket{E_2}$ and between $\ket{E_2}$ and an excited state, which then preferentially relaxes to $\ket{E_1}$ by symmetry. The steady state population of $\ket{E_1}$ is 66\%; in absence of dissipation the highest steady state population of $\ket{E_1}$ would be slightly less than 50\% (from continuously driving $\ket{0} \to \ket{E_1}$ on resonance). In (d) a similar procedure is used to stabilize a two-photon excited state $\ket{F_2}$ at higher population than would be possible from continuous two-tone driving. These ``quantum ratchet" effects are ubiquitous in the papers discussed in this review.}\label{HGfig}
\end{figure*}

An elegant experimental demonstration of state stabilization based on both energetics and symmetries was performed by Hahochen-Gourgy \textit{et al} \cite{gourgyramasesh2015} (FIG.~\ref{HGfig}), who were able to stabilize particular eigenstates in a three-qubit Bose-Hubbard model with attractive interactions. By adding a series of oscillating tones resonant with many-body transitions, the researchers were able to stabilize individual eigenstates through interaction with a lossy cavity, with fidelities between 67 and 80\%. In this case, the ratcheting effect is achieved through mixing with higher excited states that have preferential decay paths due to symmetry, rather than by incorporating additional lossy elements, but the result is similar: steady state populations of excited states which are substantially higher than what could be achieved through simple continuous driving. For example, to stabilize the 1 photon state $\ket{E_1}$ the system is resonantly driven to mix $\ket{0}$ and $\ket{E_2}$ and $\ket{E_2}$ to $\ket{F_1}$, which for symmetry reasons (the symmetric states have stronger Purcell decay into the environment than the antisymmetric ones) preferentially decays to $\ket{E_1}$ (the ratchet step, since the reverse process $\ket{E_1} \to \ket{F_1}$ does not occur), which is a comparatively long lived state. The system is thus more likely to be found in $\ket{E_1}$ than in any other state; in contrast, simply resonantly driving $\ket{0}$ to $\ket{E_1}$ produces a steady state $\ket{E_1}$ population of less than 50\%.

Following that work and a parallel scheme by Lebreuilly \textit{et al} \cite{lebreuillywouters2016}, Ma \textit{et al} \cite{maowens2017} recently proposed a mechanism for stabilizing the Mott insulating state of a strongly interacting Bose Hubbard chain. In their scheme a single auxiliary transmon qubit, which they refer to as a ``thermalizer," is added to one end of a 1d chain, and is resonantly pumped from the $\ket{0}$ to $\ket{2}$ state (the two-photon driving part of the quantum ratchet). A resonator or other lossy object is coupled to the auxiliary qubit and is chosen to be near-resonance with the $\ket{2} \to \ket{1}$ transition, giving the auxiliary qubit a very fast decay $\Gamma_{21}$ from $\ket{2} \to \ket{1}$ without significantly increasing the decay rate from $\ket{1} \to \ket{0}$ (the asymmetry in decay rates generates the ratcheting effect). If the energy of the remaining single photon is chosen to be near that of a particle in the Bose-Hubbard chain it can refill hole states created by photon losses and thus stabilize the Mott insulating state of the chain, where all qubits hold a single photon. Adding multiple auxiliary qubits could improve performance as well as potentially stabilize more complex states. Further, the refilling process these authors describe plays the role of more exotic parametric two-photon driving schemes, and could thus be applicable in more general cases. 

Subsequent work by Lebreuilly \textit{et al} \cite{lebreuillybiella2017,biellastorme2017} showed that even better performance can be obtained through a more complex, frequency-dependent bath. By considering an incoherent pumping scheme with strong frequency dependence, and in which the power spectrum decays much more quickly than $1/\omega^2$ outside the targeted region of interest, off-resonant excitations can be highly suppressed, allowing for much faster refilling rates than those permitted by the effective Lorentzian bath of a single lossy object. Such a construction could be implemented by coupling each site to many lossy objects with narrow resonances that span a precisely chosen frequency range, a structure also studied by Kapit \textit{et al} \cite{kapitchalker2015} for error correction in a topological system.

Engineered dissipation can be used to passively generate and stabilize topological states as well. Provided that the energy levels are well known and the spectrum does not include local minima that would frustrate refilling, a colored bath can stabilize anyonic states of light. One such proposal is the work of Kapit \textit{et al}, which proposes a simple mechanism to stabilize abelian and non-abelian fractional quantum Hall states \cite{kapithafezi2014}. These systems are excellent passive stabilization candidates due to the fact that the flat single particle band \cite{kapitmueller,atakisioktel} and local interactions ensure that only a single energy needs to be targeted for refilling. Schemes also exist to stabilize topological stabilizer Hamiltonians such as Kitaev's toric code \cite{kitaev} coupled to ranged interactions; these models will be discussed in the section on quantum error correction. Further discussion of many-body models with some notes on their stabilization can be found in \cite{nohangelakis2016}.

Finally, though detailed knowledge of the system's spectrum is highly desirable, such a requirement is extremely limiting. Given a desire to probe quantum dynamics of sufficient complexity to be intractable for classical computer simulation, one would like to be able to stabilize states without such knowledge, as knowing the energy levels of a system implies that it is at least partially tractable through analytic methods or classical numerics (of course, even exactly soluble models may have prohibitively complex dynamics under the right conditions). Fortunately, research in that direction is ongoing, exemplified by works such as that of Hafezi \textit{et al} \cite{hafeziadhikari2015}, or Shabani and Neven \cite{shabanineven2015}, which imagine complex dissipative structures covering a wide range of target energies at exponentially increasing rates, such that a thermal distribution is mimicked in the steady state. A rigorously benchmarked device of this type could be assumed to approximate a thermal distribution regardless of the system's many-body Hamiltonian, allowing for the quantum simulation of arbitrary strongly interacting boson models. Note however that since steady state occupation probabilities are determined from rate balancing, given the white noise photon loss model discussed earlier the lowest achievable temperature should scale logarithmically in the repair rate at the system's highest energy single photon transition divided by the photon loss rate, $\frac{E_{max}}{T_{min}} \propto \log \of{ \Gamma_{R,{\rm max}}/\Gamma_P}$. Such behavior was observed in the simulations of Kapit \textit{et al} \cite{kapithafezi2014} for a fractional Quantum Hall model with a single accessible excitation energy scale. 

\subsection{Jaynes-Cummings physics}

The works we have discussed so far have focused on using dissipation to overcome loss and stabilize quantum states or phases of matter which are ground states of the rotating frame Hamiltonian. However, continuous dissipative dynamics can create interesting many-body phases in their own right, which have no rest frame analogues. Due to their simplicity and experimental relevance, the most studied systems of this type are Jaynes-Cummings arrays of qubits and cavities, where the primary dynamics of interest occur in the cavities themselves and quantum nonlinear effects come from the dispersive qubit-cavity interaction. Depending on the drives, dissipation and cavity-cavity couplings these systems can display a range of exotic non-equilibrium effects \cite{angelakissantos2007,schmidtgerace2010,nissenschmidt2012,grujicclark2012,creatorefazio2014,rafterysadri2014,naetherquijandr2015,caomahmud2016,fossfeigniroula2016,biondiblatter2016,lebreuillywouters2016,fitzpatricksundaresan2017}, which we will very briefly review here. Qubit-cavity chains are attractive for near term experiments for many reasons, one of which being that at present it is much easier to reliably fabricate identical cavities at the same target energy than it is for qubits, where static resonance energies typically fluctuate by a few hundred MHz sample to sample even in the most mature fabrication processes.

A striking non-equilibrium feature of qubit-cavity dynamics is bistability \cite{creatorefazio2014,rafterysadri2014,caomahmud2016,fossfeigniroula2016}, where the system stochastically makes incoherent jumps between macroscopically distinct phases. This is predicted to occur at the single-cavity level \cite{drummondwalls1980} as well as in larger and more complex systems \cite{fossfeigniroula2016,fitzpatricksundaresan2017}; we will discuss an example of such bistability shortly, where a chain of cavities stochastically flips between states with transmission coefficients differing by multiple orders of magnitude. Another interesting phenomenon in clusters of qubits and cavities is self-trapping, as explored in \cite{schmidtgerace2010,rafterysadri2014}. We consider a pair of cavities coupled by a simple exchange interaction $J \of{a_L^\dagger a_R + a_R^\dagger a_L}$, and each coupled to a resonant qubit with an exchange interaction of strength $g$; for simplicity we will take the qubit to be a true two-level system. If the system is initialized with an imbalanced population for $g=0$ the imbalance will simply oscillate between the two cavities with frequency $2J$, but at finite $g$ the oscillations become anharmonic, and at a particular critical coupling $g_c \propto J \sqrt{N}$ (where $N$ is the number of photons) the oscillation frequency vanishes, localizing the photon imbalance. With continuous driving and dissipation the situation is considerably more complex, as non-equilibrium effects delocalize the system at sufficiently long time; if the system is initialized with a number of photons $N > N_c$ (the critical number for self trapping) the system will cross a phase transition and localize as the photon population decays. Near this transition point critical effects cause superexponential decay in the measured homodyne microwave signal, and once localized phase coherence in the signal rapidly collapses.

\begin{figure}
\includegraphics[width=3.5in]{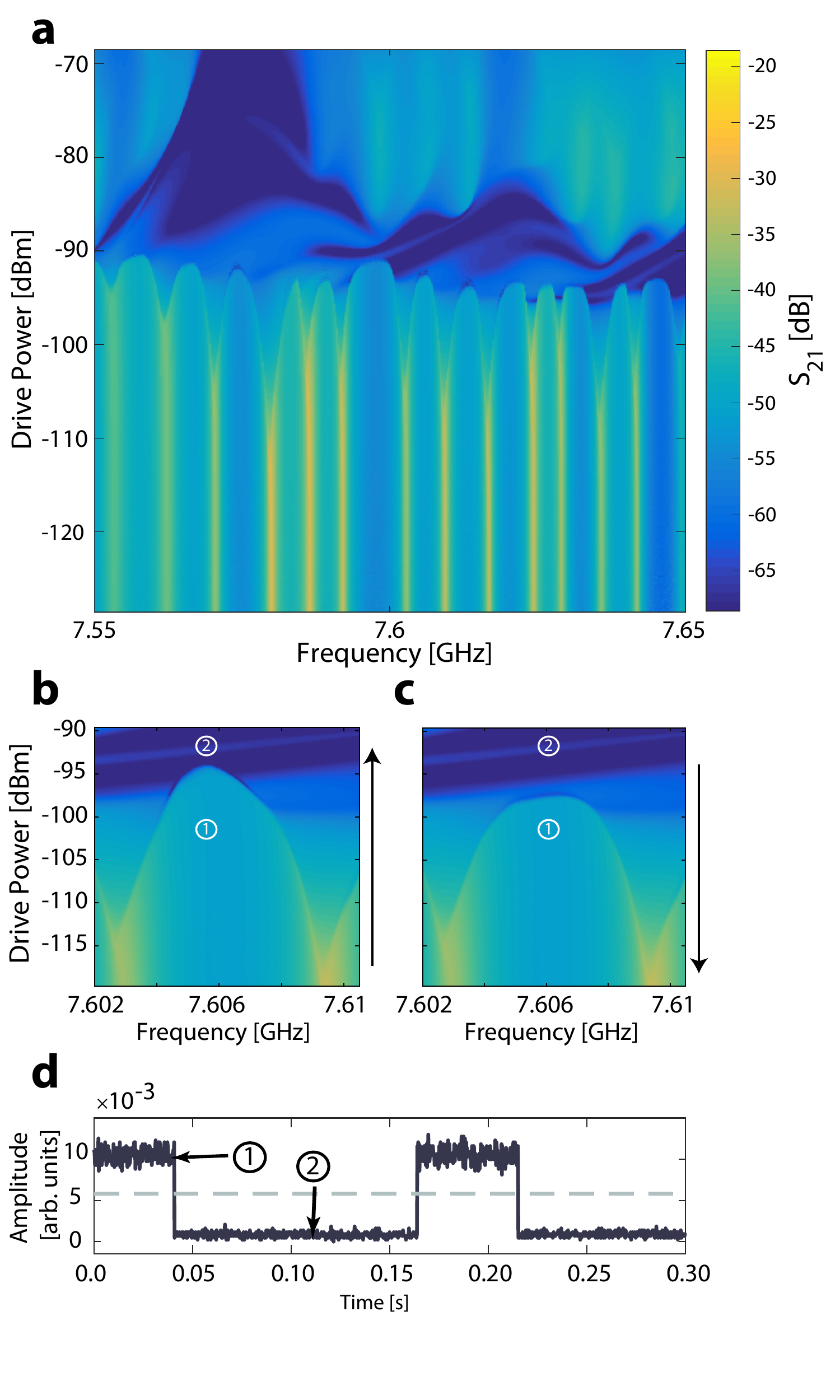}
\caption{Phase transitions and bistability in a chain of 72 qubit-cavity pairs, reproduced with permission from \cite{fitzpatricksundaresan2017}. In (a) the transmission coefficient $S_{21}$ along the chain is plotted as a function of frequency and input power. At low power, $S_{21}$ has peaks and valleys consistent with single particle resonances of tunneling modes along the chain, but at high power quantum many-body effects cause a transition into a region of strong scattering where $S_{21}$ drops by multiple orders of magnitude. This transition exhibits hysteresis and bistability; in (b) and (c) the same window of drive power and frequency is examined. For each frequency the system is ramped from low power to high (b) or high power to low (c) over 31.95ms, and the differing locations of the transition indicate hysteresis. In (d), a single trace at constant power is presented, tuned to lie very close to the transition point. The spontaneous jumps between transmitting and insulating phases (1) and (2), captured here in a single shot trace of the homodyne phase, demonstrate bistability over periods of tens of ms, far longer than any other timescale in the system.}\label{fitzpatrickfig}
\end{figure}

The largest experiment so far was performed by Fitzpatrick \textit{et al} \cite{fitzpatricksundaresan2017}, and consisted of 72 qubit-cavity pairs in a linear chain, well into the many-body regime. As shown in FIG~\ref{fitzpatrickfig}, the experimenters measured the transmission $S_{21}$ along the chain as a function of input frequency and applied drive power. In the weak driving limit, $S_{21}$ displays a series of frequency peaks consistent with the single particle energies of a photon hopping along the chain, as we would expect from a few-body quantum system where the density is low enough that the particles are nearly free. 

In contrast, as the power is increased the system exhibits a sharp transition into an ``insulating" regime where the transmission coefficient drops by more than two orders of magnitude. The authors attribute this effect to $N$-photon resonances between the cavities and the transmons themselves; though the bare cavity frequency of $\omega_c \sim 2\pi \times 7.6$ GHz is far less than the transmon energy $\omega_T \sim 2 \pi \times 12.7-13.9$ GHz, the transmon nonlinearity is negative and grows more negative as the energy increases, so at large enough $N$ resonant exchange is possible between the cavities and the qubits. When this occurs the transmons can act as strong impurity scatterers that strongly suppress transmission. 

More remarkably, the many-body ``insulating" phase exhibits bistability and hysteresis near the transmission regions, as predicted for smaller systems \cite{caomahmud2016,fossfeigniroula2016,biondiblatter2016}. As shown in the figure, this switching occurs stochastically and very rapidly, though the two phases are themselves metastable with average switching times on the order of tens of ms, which are four orders of magnitude larger than the photon lifetimes of the qubits and cavities themselves. These states also display hysteresis, with the transition occurring at different drive powers depending on whether the experiment begins at high power and ramps down or begins at low power and slowly ramps up. Due to the vastly increased complexity the authors were unable to predict many of these effects quantitatively using mean-field simulations, suggesting that interesting quantum effects were responsible. Undoubtedly more surprises would be in store were the system extended to two dimensions, and by combining a 2d array with time reversal breaking hopping terms, one could realize fractional quantum Hall states of light \cite{carusottociuti2013,kapithafezi2014,andersonma2016,roushanneill2017}.

\subsection{Complexity}

We shall conclude this section with a note about quantum complexity. An important near-term goal of the field is to demonstrate ``quantum supremacy"-- that is, engineered quantum dynamics in a collection of qubits which are beyond the scope of classical simulation \cite{preskill2011}. For example, Boixo \textit{et al} \cite{boixoisakov2016} consider the problem of sampling the output of random quantum gate sequences. Though not useful for any practical concern, showing that this problem is classically hard is easy, as if it were not, all quantum computing algorithms could be efficiently simulated on a classical computer. The computationally challenging part of simulating a particular gate sequence is sparse matrix-vector multiplication, so the total cost in both time and memory is a polynomial in $N$ times the size of the Hilbert space ($2^N$). Simulating unitary Hamiltonian time evolution has similar computational scaling. Boixo \textit{et al} report than a 42 qubit simulation stretches the limits of modern supercomputers, and moving to 49 qubits (a simple $7 \times 7$ grid) would leave the realm of classical simulation entirely. Further, one need not only consider qubits; arrays of linear quantum optical elements can also be extremely difficult to simulate, a problem known as Boson Sampling \cite{aaronsonarkhipov2013}.

The dynamics of driven-dissipative many-body quantum systems are complex (in both the intuitive and computational meanings of that word), potentially much more so than unitary time evolution. These dynamics are often poorly understood, due to both the large parameter space of possible models and the computational complexity of simulating them. Now, if we assume a simple Markovian white noise model for dissipation, as is appropriate in most superconducting device experiments, a system's exact quantum dynamics can be simulated through the Lindblad master equation \cite{gardinerzoller} or quantum jump methods (also referred to as the ``Monte Carlo Wavefunction" method), reviewed by Plenio and Knight \cite{plenioknight1998} and more recently by Daley  \cite{daley2014}. In quantum jump calculations the system's wavefunction undergoes ordinary Hamiltonian evolution and dissipation is captured by applying random instantaneous jumps, with physical observables computed by averaging over many random trajectories, at a total cost which is only polynomially more expensive than direct Hamiltonian evolution in closed systems. And as with closed systems, in many cases the system's Hamiltonian may have symmetries or be amenable to analytical and approximate treatments that can further simplify the problem. 

However, if the noise is non-Markovian, or even Markovian with relative rates that depend on the energetics of the Hamiltonian (the ``colored baths" that appear ubiquitously in this review) the problem difficulty can increase enormously. In the case of colored baths, if the transition rate between two eigenstates depends on their relative energies and not just the matrix element from a photon loss operator then the system can only be accurately simulated if those energies are known. For general systems this requires exact diagonalization, and if these energies are far from the ground state Lanczos methods are not applicable, so the cost to compute all of them scales as $2^{2N}$ in memory (and even worse in time). Thus dissipative systems can present significantly greater computational challenges than closed ones, though a variety of techniques have been developed to study then-- see \cite{albertbradlyn2016,lipetruccione2016} and references therein. 

To be more concrete, consider a cluster of $N$ qubits, driven by oscillating fields, and coupled to each other through tunable coupling elements. Each qubit is coupled to a readout resonator, with multi-photon processes used to engineer driven loss and/or refilling through near-resonant photon exchange and/or parametric two photon creation and annihilation. The qubits and resonators lose photons at rates $\Gamma_{iP}$ and $\Gamma_{iS}$, which are fixed when the experiment is designed. The total rotating frame system Hamiltonian is:
\begin{eqnarray}
H &=& \sum_{ij} \hat{O}_{ij} + \sum_i \of{ \overrightarrow{h}_i \cdot \sigma_{iP} } + \sum_i \omega_{iS} a_{iS}^\dagger a_{iS} \\
& & + \sum_i  \of{ \Omega_{i}^+ \sigma_{iP}^+ a_{iS}^\dagger + \Omega_{i}^- \sigma_{iP}^+ a_{iS}  + {\rm H.c.} }. \nonumber
\end{eqnarray}
Here, the $\hat{O}_{ij}$ are two-qubit operators, which could be simple exchange couplings $J_{ij} \of{\sigma_{iP}^+ \sigma_{jP}^- + {\rm H.c.} }$. All the terms in $H$ are tunable (and may be varied during the system's evolution), with only the loss rates fixed. The computational task to simulate could be to simply sample the output probabilities in the $z$ basis of the qubits (using the resonators to measure them), either at a finite time given some known initial condition, or in a long time steady state. Given appreciable $\Omega_{i}^\pm$ the internal dynamics of the resonators cannot be ignored and the simulation cost rises to at least $2^{2N}$. This cost could grow higher still, depending on the mixing with two-photon states in qubits and resonators. 

One might object that it is possible for the lossy resonators to permit easier classical simulation of this system, either if the noise allows new classical algorithms to be used, or if the qubit output probabilities become sufficiently uninteresting, e.g. the empty state or classical randomness. Given the examples elsewhere in the review I find both of these outcomes exceedingly unlikely in the general case; if the qubit-qubit interaction energy scales are large compared to the $\Omega_{i}^{\pm}$ the excited state of a given resonator can easily become entangled with other qubits some distance away, and the resulting steady states of the qubits can have plenty of structure depending on how the various terms are chosen. The experimental challenge, then, could be to compare the output distribution observed to a numerical simulation given the input parameters in $H$; uncertainty in the parameters' physical values, measurement inaccuracy, and (potentially) $1/f$ phase noise would all have to be well-controlled to ensure the distributions match. Given relatively long $T_1$ so that $\Gamma_{iP}$ is much smaller than any of the incoherent transition rates induced by the qubit-resonator couplings one could even compare the results of the physical evolution to a simulation where $\Gamma_{iP}$ is set to zero. However, the feasibility of obtaining accurate simulations of the infinite $T_1$ state (either at long time or at finite time given known initial conditions) even at relatively small $N$ would require complex calculations far beyond the scope of this review.

We conclude that the dynamics and steady states of driven-dissipative systems evolving under a many-body Hamiltonian and subject to colored noise can be significantly more expensive to simulate than Hamiltonian time evolution. This in turn means that quantum dynamics beyond the scope of classical simulation could be achieved with a smaller number of qubits, which could themselves be noisy as it is noise that makes the system difficult to study! That said, to establish quantum supremacy one would need to construct certification criteria to verify that the desired driven-dissipative dynamics have been achieved; in Boixo \textit{et al} quantum dynamics is distinguished from classical randomness by comparing the result to the Porter-Thomas distribution expected for random quantum circuits, with fidelity determined from cross-entropy. Adapting such measures to a driven-dissipative system would undoubtedly require subtle considerations beyond the scope of this review; I include this section merely to suggest that driven-dissipative systems could be a fertile ground for near-term research in quantum supremacy. 

\section{Quantum error correction}

Perhaps the most important application of passive state stabilization is quantum error correction. The sufficient conditions for passive QEC in an engineered primary system and bath are (i) that the rotating frame ``ground" states of the primary system form a degenerate ``logical" manifold, (ii) the dissipation-induced repair rate for excitations created by photon losses is much faster than the primary loss rate itself, which generally implies (iii) up to small corrections, the long-time steady state of the system should be given by an incoherent mixture of states in the logical manifold, and most importantly, (iv) that the total process of a photon loss and its subsequent correction produces no relative phase, rate or energetic differences between orthogonal states in the logical manifold. Achieving all these requirements in a system simple enough for experimental implementation is a subtle problem, and I will review two prominent examples (``Cat codes" and the ``Very small logical qubit") in this section. But before delving into these architectures, the astute reader may note that the conditions I have just declared include only a single quantum error channel, photon loss, and ignore the other two, photon addition and dephasing. As discussed earlier, photon addition can be largely ignored on energetic grounds, as the typical resonance energy (3-10 GHz) of photons in superconducting circuits is much larger than the ambient temperature (15-50 mK), and the resulting population of thermal photons at the resonance energy is exponentially suppressed \cite{jinkamal2015}. Dephasing, however, is a more subtle issue that we will now consider.

Let us first imagine we have engineered a primary system Hamiltonian $H_P$, with degenerate ``logical" ground states $\ket{0_L}$ and $\ket{1_L}$, separated from all other states by an energy of at least $\Delta$. Since a photon loss is a non-unitary operation that combines an excitation removal with a measurement, satisfying condition (iv) requires that
\begin{eqnarray}
\bra{0_L} a_i^\dagger a_i \ket{0_L} = \bra{1_L} a_i^\dagger a_i \ket{1_L}, \; \bra{1_L} a_i^\dagger a_i \ket{0_L} = 0, 
\end{eqnarray}
holds, up to very small corrections, for all degrees of freedom $i$ in the system. Since phase noise acts through the number operator $a_i^\dagger a_i$, the only way it can mix logical states is by exciting the system into states outside of the logical state manifold, at a finite energy cost $\geq \Delta$. 

If this occurred frequently, it would be a significant error source that would need to be compensated with further bath engineering, but fortunately the action of the primary Hamiltonian $H_P$ ensures that the system is only sensitive phase noise near the transition frequency between these states. Since dephasing in superconducting circuits has a low-frequency dominated power spectrum $S \of{\omega} \propto 1 / \omega^\alpha$ (with $\alpha \simeq 1$), if we only sample the noise at $S \of{\Delta}$ its effect will be dramatically weakened compared to the free induction decay case of $H_P =0$. Thus, just as continuous Rabi driving extends the coherence of single qubits against phase noise \cite{martinisnam2003,yoshiharaharrabi2006,bylandergustavsson2011,yangustavsson2013,paladinogalperin2014,omalleykelly2015}, the action of any $H_P$ satisfying the above conditions for photon loss error correction will not commute with local $a_i^\dagger a_i$ and strongly suppress errors from phase noise as a result (this also applies to quantum simulation, as demonstrated indirectly in \cite{roushanneill2017}). Empirically, given a single qubit Ramsey $T_{2R}$ as a measure of the $1/f$ phase noise strength, this suppression produces an effective logical error time from phase noise of $T_{2L} \propto \Delta \of{ T_{2R}}^2$, yielding quadratic improvements in phase error time from a linear improvement in $T_{2R}$. Phase noise suppression is therefore an intrinsic benefit of nearly any continuously applied Hamiltonian \footnote{If $H_P$ generated by a digital sequence of pulses the same suppression is not similarly guaranteed, since phase errors are free to accumulate between application of the various terms in $H_P$. However, if these free induction decay times are short or frequently interrupted low frequency phase noise will still be weakened by the system's rapid ``toggling," as in spin echo sequences.} capable of correcting photon loss errors, massively simplifying the circuit's design! Note of course that this does not hold in systems experiencing high-frequency phase noise, but fortunately this is not a generic situation for driven superconducting circuits.

\subsection{Bit-flip code}

\begin{figure}
\includegraphics[width=3.0in]{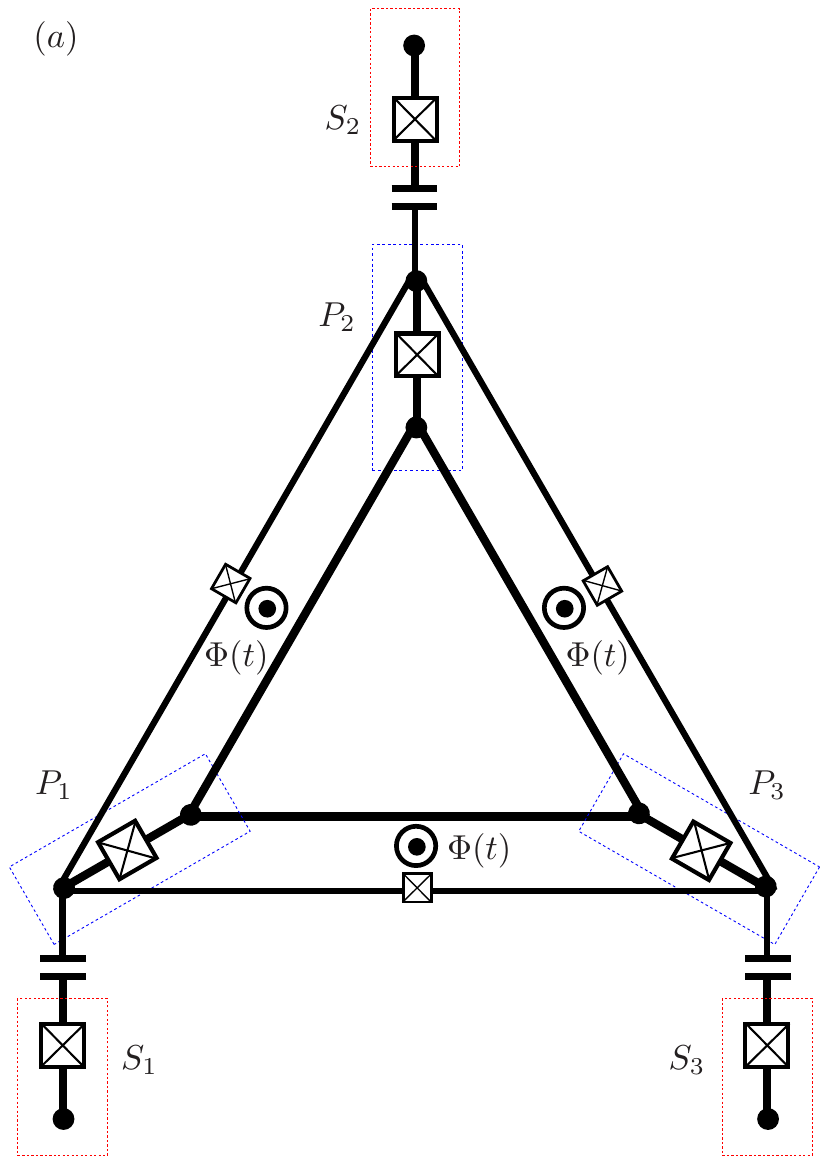}
\includegraphics[width=3.0in]{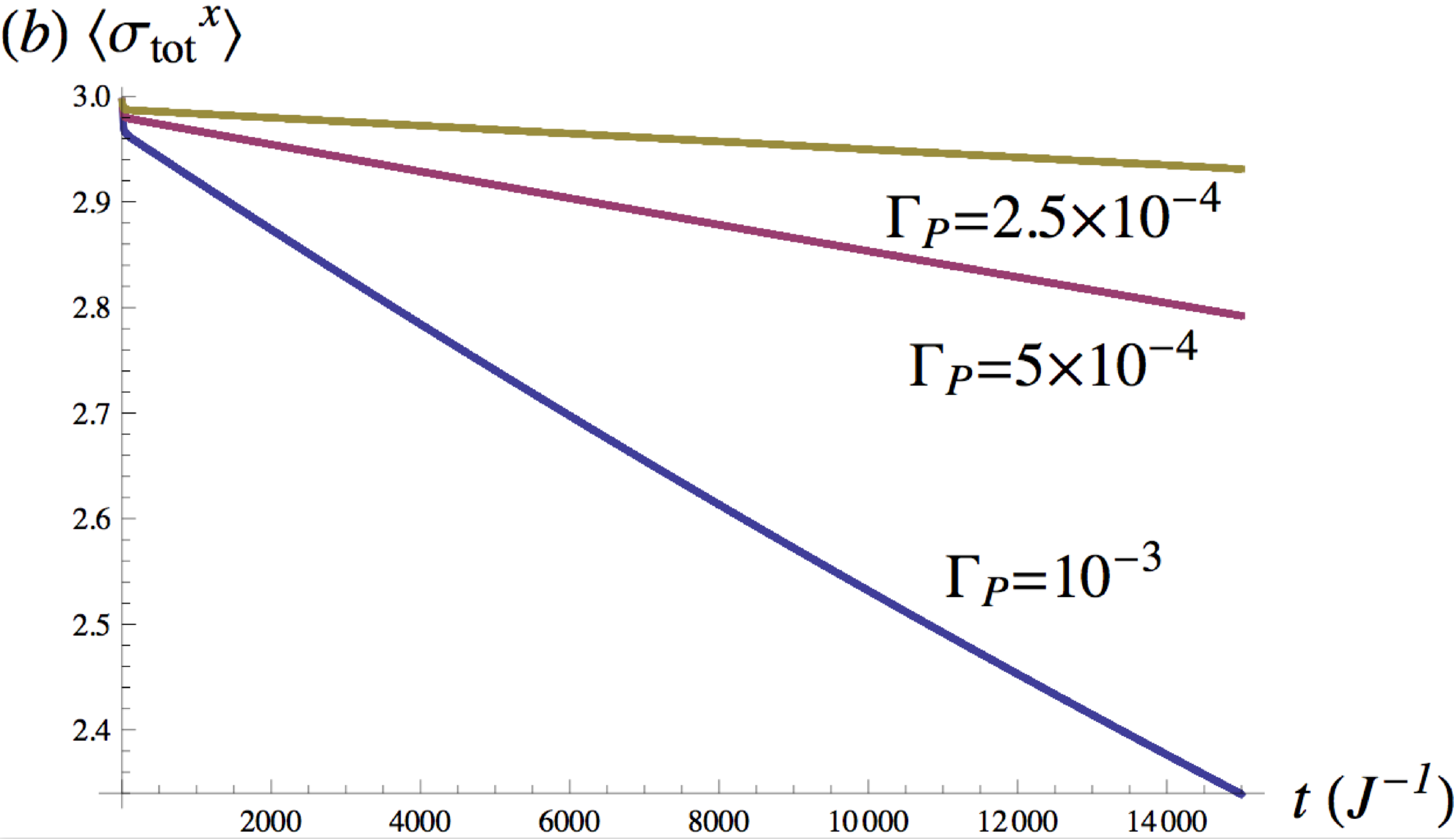}
\caption{(Color online) A simple circuit to demonstrate passive error correction, taken from \cite{kapithafezi2014}. (a) A ring of three primary transmon qubits ($P_1,P_2,P_3$, blue) are coupled to each other through flux biased Josephson junctions, which are driven at the sums and differences of the qubit frequencies \cite{kapit2015}. These terms produce an effective interaction $- J \sigma_{iP}^{x} \sigma_{jP}^{x}$ in the rotating frame. Each primary qubit is weakly coupled capacitively to a lossy shadow qubit ($S_1,S_2,S_3$, red; these objects could also be resonators as the nonlinearity is unimportant). In (b), the system is initialized in the ground state $\ket{111}$ in the $x$ basis, and the Lindblad equation is numerically integrated to track the system's evolution. Parameters used are $\Omega = 0.05 J$ and $\Gamma_S = 0.1 J$ ($\Gamma_R = 0.05 J$), and three primary qubit decay rates $\Gamma_P = \cuof{10^{-3}, 5 \times 10^{-4}, 2.5 \times 10^{-4}} J$. The resulting logical error rates $\Gamma_L$ (see Eq.~\ref{3qubitGL}) are $\cuof{4.6 \times 10^{-5}, 1.3 \times 10^{-5}, 3.8 \times 10^{-6}} J$, demonstrating near-quadratic increases in logical state lifetime from linear decreases in $\Gamma_P$.}\label{ringfig}
\end{figure}

As a warmup to more complex models, let us consider a driven-dissipative implementation of the three-qubit bit flip code \cite{reeddicarlo2}, proposed independently by Cohen and Mirrahimi \cite{cohenmirrahimi2014}, and Kapit, Hafezi and Simon \cite{kapithafezi2014}. A passive three qubit bit flip code was also considered in a different context by Kirchoff \textit{et al} \cite{kerckhoffnurdin2010} in 2010. Using the language of \cite{kapithafezi2014}, the Hamiltonian consists of six qubits, three of which are high coherence (with a low loss rate $\Gamma_P$) and form the primary $H_P$ and three of which are lossy ``shadow" qubits (with fast loss rate $\Gamma_S$) used as an engineered reservoir. The system Hamiltonian, engineered through a combination of fixed Josephson and capacitive couplings and tones at multiple frequencies, reduces to
\begin{eqnarray}
H_P &=& - J \of{\sigma_{1P}^x \sigma_{2P}^x + \sigma_{2P}^x \sigma_{3P}^x + \sigma_{1P}^x \sigma_{3P}^x }, \\
H_{S} &=& 2 J \sum_{i=1}^3 \sigma_{iS}^z, \; H_{PS} = \Omega \sum_{i=1}^3 \of{\sigma_{iP}^x \sigma_{iS}^x + \sigma_{iP}^y \sigma_{iS}^y}, \nonumber \\
H &=& H_P + H_S + H_{PS}. \nonumber
\end{eqnarray} 
Note that the device as laid out above does not satisfy condition (iv) of the previous section. This is because a photon loss in a primary qubit, through the $\sigma_{iP}^- = \sigma_{iP}^x - i \sigma_{iP}^y$ operator, has a 50\% chance of enacting a $\sigma^x$ operation, which does not take the system out of its degenerate ground state but does dephase a superposition within that manifold. Nonetheless, it provides a useful toy model to understand passive quantum error correction. This Hamiltonian could be implemented through flux driven Josephson couplers as described in \cite{kapit2015} (see FIG.~\ref{ringfig}), or through capacitively coupled qubits subject to multiple drives \cite{cohenmirrahimi2014}, and it has a twofold degenerate ground state with all shadow qubits in $\ket{0}$ and the three primary qubits in logical states $\ket{0_L} = \ket{000}$ or $\ket{1_L} = \ket{111}$ in the $x$-basis.

The simplicity of this system allows us to predict the transition rates $\Gamma_{ij}$ induced in the primary system by the shadow qubits using Fermi's golden rule. The dominant contribution through this is through the $yy$ part of the capacitive interaction, which flips the state of a primary and a shadow qubit simultaneously. Taking into account that the shadow qubit must relax back to its ground state (at a rate $\Gamma_S$) to complete the process, and assuming simple Lorentzian line broadening for the shadow qubit, we can compute the total rate $\Gamma_{ij} \of{\delta E}$ for a process which changes the primary qubits' energy by $\delta E$,
\begin{eqnarray}
\Gamma_{ij} \of{\delta E } &=& \of{ \Gamma_{PS}^{-1} \of{\delta E} + \Gamma_S^{-1} }^{-1}, \\
&=& \frac{ \Omega^{2} \Gamma_S  }{ \Omega^2 + \of{\delta E + 4 J}^{2} + \Gamma_S^{2} /4    }. \nonumber
\end{eqnarray} 
When applied to the error correction code above, we find a repair rate $\Gamma_R$ for removing a phase flip ($\delta E = - 4J$) and relaxing the system back to its ground state along with two induced error rates, $\Gamma_{E0}$ corresponding to off-resonantly flipping a second qubit after one flip has already occurred (at zero energy cost in the primary system) and $\Gamma_{E1}$ corresponding to an off-resonant phase flip when the primary system is in a logical state. Combining these three rates and assuming an intrinsic (white noise) $y$ error rate of $\Gamma_P$ in the primary qubits, we conclude:
\begin{eqnarray}\label{3qubitGL}
\Gamma_R &=& \frac{\Omega^2 \Gamma_S}{\Omega^2 + \Gamma_S^2 / 4 }, \; \Gamma_{E0} \simeq \frac{\Omega^2 \Gamma_S}{16 J^{2}}, \; \Gamma_{E1} \simeq \frac{\Omega^2 \Gamma_S}{64 J^{2}}, \\
\Gamma_L &\simeq& 6 \of{ \Gamma_P + \Gamma_{E1} } \frac{ \Gamma_P +  \Gamma_{E0}  }{ \Gamma_P + \Gamma_{E0} + \Gamma_R } \simeq 6 \frac{\Gamma_P^2}{\Gamma_R}. \nonumber
\end{eqnarray}
This decreases quadratically with linear increases in $\Gamma_P$ and thus acts as quantum error correction. Note that, as seen in FIG.~\ref{ringfig} at short times $t \leq 1/\Omega$ the passive error correction provides no protection and the transient error rate is $\approx 3 \Gamma_P$. This represents the transient probability of catching the system between a $y$ error and it's subsequent correction by a shadow qubit, and has no effect on the long time rate of decay. It can however be an important challenge for designing error-tolerant gates for these qubits, a subject which is beyond the scope of this review. 

If we further consider Eq.~(\ref{3qubitGL}), we notice also that minimizing $\Gamma_L$ at fixed $\Gamma_P$ and $J$ suggests the hierarchy of scales
\begin{eqnarray}\label{ECcond}
J/\hbar  \gg \Omega \sim \Gamma_S \gg \Gamma_P.
\end{eqnarray}
This order of inequalities is a generic feature of any passive error correction mechanism where error states are differentiated from logical states on energetic grounds, including the substantially more complex topological schemes discussed below. This is the heart of error correction through engineered dissipation: error syndrome operators are implemented to induce an energy penalty for error states, which can then be corrected through tuned couplings and energetic resonance with a lossy subsystem. And fortunately, the relatively high coherence of superconducting qubits makes (\ref{ECcond}) easy to satisfy even with current technology. If we assume our primary system to be a driven multi-qubit device then the many-body energy scale $J$ is set by the strength of qubit-qubit couplings, which in modern designs are typically in the tens of MHz (times $2 \pi$). In contrast, for a typical transmon qubit $\Gamma_P \sim 20 - 100$ kHz, leading to $J/\Gamma_P \geq 10^{3}$. This wide ratio leaves substantial room for $\Omega$ and $\Gamma_S$ (and thus $\Gamma_R$) in the range of a few MHz, which would already be faster than the fastest digital error correction schemes demonstrated in the literature \cite{kellybarends2015}, which have a total cycle time of around 1 $\mu$s.

\subsection{Cat codes}

Cat codes \cite{leghtaskirchmair2013,mirrahimileghtas2014,sunpetrenko2014,leghtastouzard2015,albertshu2016,ofekpetrenko2016,cohensmith2016,mundhadagrimm2016,michaelsilveri2016,wanggao2016,heeresreinhold2016,puriblais2016}, where quantum information is encoded in superpositions of coherent states in 3d superconducting cavities, have achieved dramatic success in recent years, culminating in the first demonstration of quantum error correction (in any platform) which extends the lifetime of a state beyond its most coherent part \cite{ofekpetrenko2016}. They are part of a broad class of continuous variable codes, with historical roots in the Gottesman-Kitaev-Preskill codes \cite{gottesmankitaev2001}, and operate in a regime where the many-photon state of driven resonator is specified by its coherent phase $\alpha$. In this section, I review the basics of the cat code and provide an overview of current research on them. As we shall see, though engineered dissipation is not required to implement a cat code in its most basic form, it can substantially improve performance and will be a necessary part of future developments in these devices.

\begin{figure}
\includegraphics[width=3.5in]{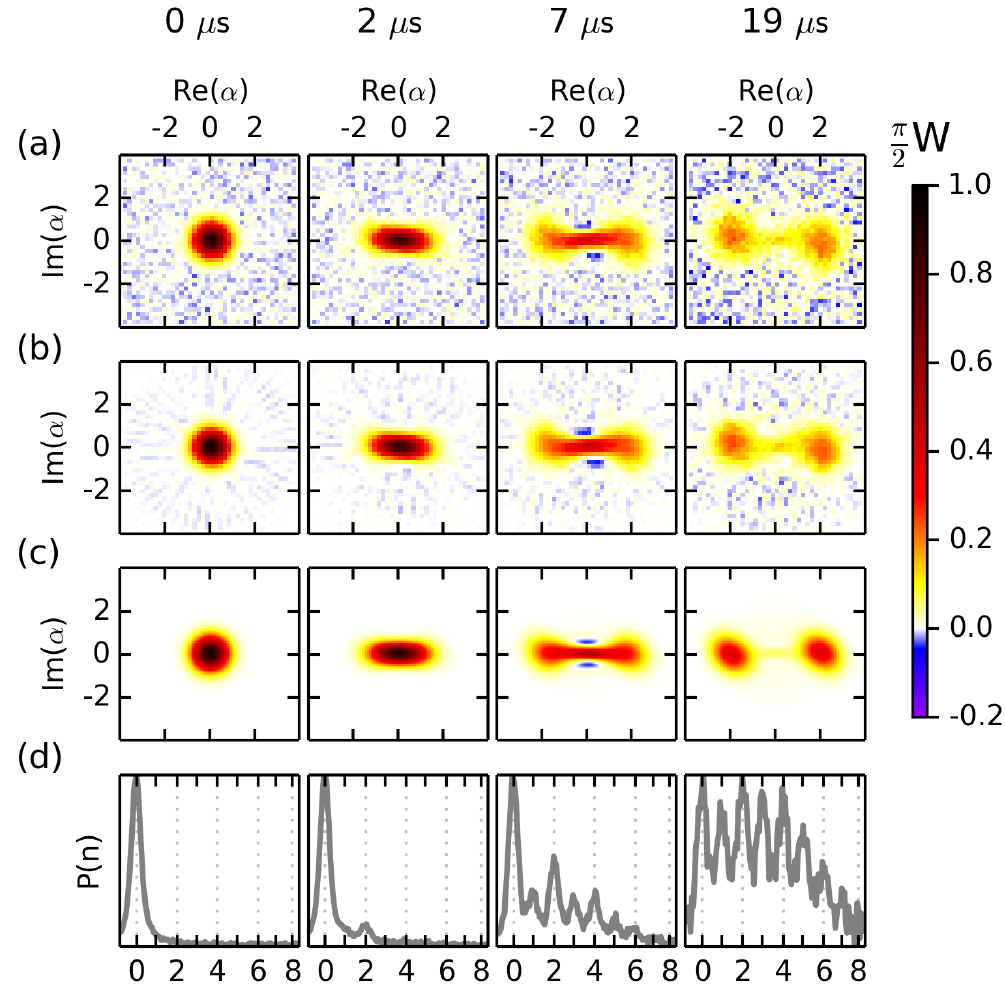}
\caption{Evolution of a cat state under two-photon driving and dissipation, reproduced with permission from \cite{leghtastouzard2015}. The system begins empty of photons and at $t=0$ two-photon driving and dissipation are turned on, stabilizing a two-photon cat state $\ket{\alpha} + \ket{-\alpha}$. Rows (a) and (b) represent experimental data, row (c) is a theoretical simulation and row (d) is the probability $P_n$ of occupying each $n$ photon state, as a function of time. Due to much slower single photon loss the system eventually decays to an incoherent mixture of states. These two photon states suppress dephasing but are mixed by a photon loss; a manifold of four-photon states can be used to protect quantum information against all single photon error channels.}\label{cat2fig}
\end{figure}

The cat code in its simplest form consists of two resonators (one of which is high coherence and the other which is lossy and used for readout) and a transmon qubit, which is used to manipulate the state of the primary resonator. Specifically, it takes as its basis the coherent states
\begin{eqnarray}
\ket{\alpha} \equiv  e^{-\abs{\alpha}^2 / 2} \sum_{n=0}^{\infty} \frac{\alpha^n}{\sqrt{n!}} \ket{n}.
\end{eqnarray}
To use these states for quantum error correction, we need to construct a reduced subspace of two or more states. Before presenting the full cat code, we shall discuss a simplified version which protects only against phase noise. We consider a high coherence cavity $P$ subject to two-photon driving and dissipation as demonstrated in \cite{leghtastouzard2015}, with rotating frame Hamiltonian $H = \Omega \of{a_P^\dagger a_P^\dagger + a_P a_P}$. This Hamiltonian is engineered through dispersively coupling the high coherence cavity to a qubit which is then coupled to a lossy readout cavity $R$, and driving the qubit $Q$ at $2 \omega_P - \omega_R$ and much more weakly at $\omega_R$. The net result of this nonlinear process is to convert pairs of photons from the primary resonator into a photon in the readout resonator (which decays rapidly). This produces both coherent and incoherent two-photon terms in the primary resonator: the resulting effective Hamiltonian is $H \simeq \Omega \of{a_P^\dagger a_P^\dagger + a_P a_P}$, and along with it come Lindblad collapse operators $\sqrt{\Gamma_P} a_P$ and $\sqrt{\Gamma'} a_P a_P$, where $\Gamma_P$ is the damping rate of the primary cavity and $\Gamma'$ is the induced two-photon loss through coupling to the driven, lossy readout resonator. For a sufficiently high coherence cavity $\Gamma' \gg \Gamma_P$ the steady state of the system is a mixture of $\ket{0_L} \simeq 2^{-1/2} \of{ \ket{\alpha} + \ket{- \alpha}}$ and $\ket{1_L} \simeq 2^{-1/2} \of{ \ket{\alpha} - \ket{- \alpha}}$ (where $\alpha$ is determined by the drive strength and dissipation rates). Since the orthogonality between $\ket{\alpha}$ and $\ket{-\alpha}$ is nonzero (though exponentially small in increasing $\abs{\alpha}$) the $1/\sqrt{2}$ normalization is not exact. As shown in FIG.~\ref{cat2fig}, starting from all drives turned off with the system unpopulated, the pair of cavities will evolve deterministically to $\ket{0_L}$ or $\ket{1_L}$ depending on the relative phase of the drives, before decaying into an incoherent mixture of the two states at long times.

These states can be distinguished by the parity operator $e^{i \pi a_P^\dagger a_P}$, which can be measured by dispersive coupling to an off-resonant qubit. Further, they are protected against low frequency phase noise in the primary cavity. Specifically, if we consider phase noise through the number operator $a_P^\dagger a_P$, then we find both $\bra{0} a_P^\dagger a_P \ket{1}$ and $\bra{1} a_P^\dagger a_P \ket{1} - \bra{0} a_P^\dagger a_P \ket{0} \propto e^{-\abs{\alpha}}$, so phase noise cannot effectively mix the two logical states. Phase noise can cause the state to drift away from the $\ket{\pm \alpha}$ basis (and out of our logical state manifold), but this is energetically suppressed by the continuous driving as discussed earlier. Photon losses, however, will swap the two states, with $\bra{0} a_P \ket{1} \simeq \alpha$. Consequently, a more complex scheme is needed to protect against all relevant error channels.

\begin{figure}
\includegraphics[width=3.5in]{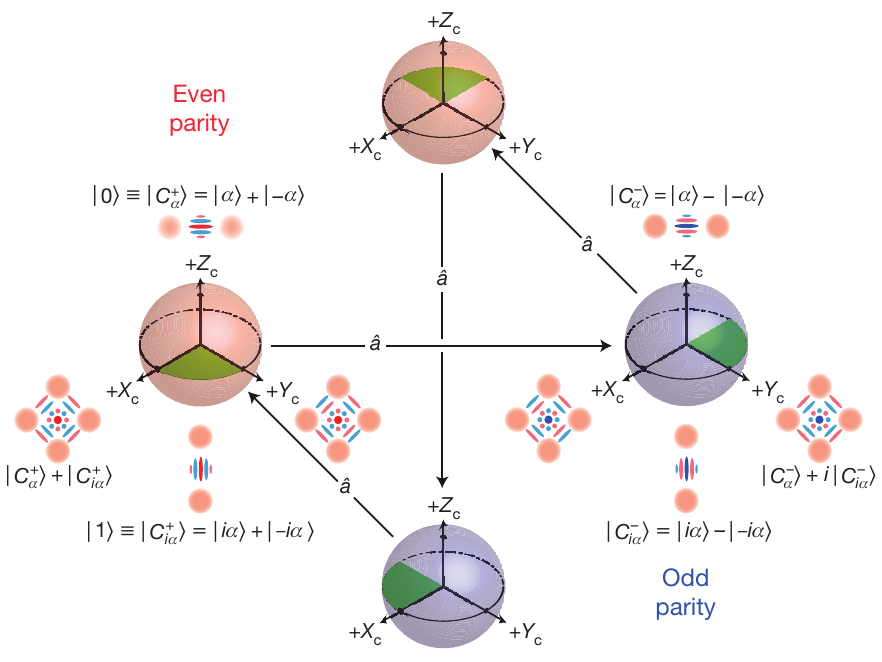}
\caption{Cat code state space, reproduced with permission from \cite{ofekpetrenko2016}. Logical states are constructed from superpositions containing an even number of photons; they are spanned by the manifold in Eq.~(\ref{catstates}). When a photon is lost from the primary resonator the logical state manifold is mapped to the error manifold (\ref{errstates}), and since the states in the error manifold have unique parent states in the logical manifold, with no phase or amplitude differences that depend on the parent states (up to factors that scale as $e^{-\abs{\alpha}^2}$, where $\abs{\alpha}$ is the resonator amplitude), no information is lost and no unwanted gates are enacted. The setup is thus suitable for quantum error correction, as described in the text.}\label{catstatesfig}
\end{figure}

\begin{figure*}
\includegraphics[width=6in]{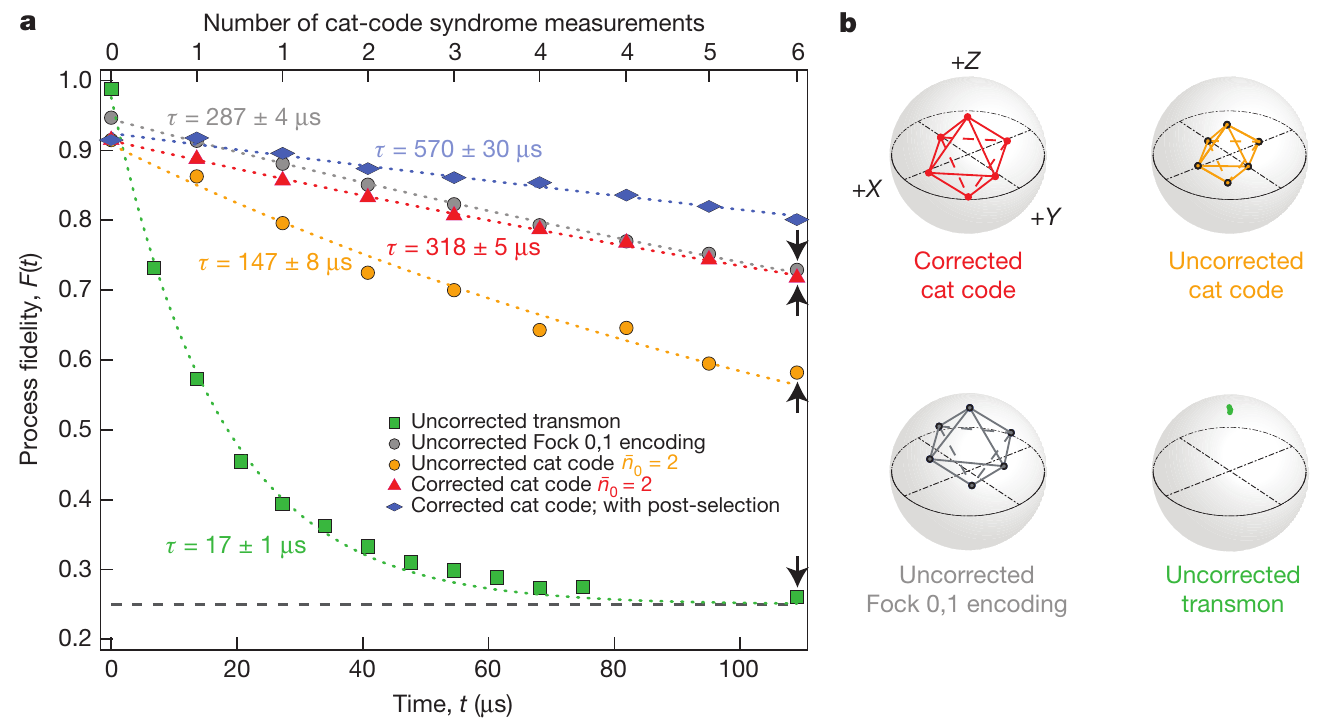}
\caption{Extending the lifetime of a logical state through cat code error correction, reproduced with permission from \cite{ofekpetrenko2016}. Repeated photon parity measurements track photon losses in a primary resonator without distinguishing between logical states, creating an error detection code that preserves quantum information. In (a) the lifetime of encoded states is shown; the gray Fock state curve is the lifetime of a single photon in the cavity (the most coherent degree of freedom in the circuit). The orange and red curves represent encoded cat states as in Eq.~(\ref{catstates}), with and without repeated parity detection to track errors. The resulting lifetime of the error-corrected states is about 10\% higher than that of the Fock states, demonstrating useful quantum error correction that exceeds break-even. Note that the cat states in this experiment eventually decay to vacuum as no mechanism exists to replace lost photons; possible mechanisms for photon refilling are discussed in the text.}\label{catECfig}
\end{figure*}

The minimal cat code capable of protecting against both losses and dephasing requires coherent four-photon driving and dissipation \cite{mirrahimileghtas2014}. The implementation of such a drive configuration is technically challenging, but it can be achieved through chaining two-photon drives in sequence \cite{mundhadagrimm2016}. To demonstrate error correction, let us consider the basis of states spanned by $\ket{\pm \alpha}$ and $\ket{\pm i \alpha}$. We consider the superpositions shown in FIG.~\ref{catstatesfig}:
\begin{eqnarray}\label{catstates}
\ket{0_L} &\equiv& \frac{\ket{\alpha} + \ket{- \alpha} + \ket{i \alpha} + \ket{-i \alpha}}{2} \\
&=& 2 e^{-\abs{\alpha}^2 / 2} \sum_{n = 0}^{\infty} \frac{\alpha^{4n}}{\sqrt{4n!} } \ket{4n}, \nonumber \\
\ket{1_L} &\equiv& \frac{\ket{\alpha} + \ket{- \alpha} - \ket{i \alpha} - \ket{-i \alpha}}{2} \nonumber \\
&=& 2 e^{-\abs{\alpha}^2 / 2} \sum_{n = 0}^{\infty} \frac{\alpha^{4n+2}}{\sqrt{\of{4n+2}!} } \ket{4n+2}. \nonumber
\end{eqnarray}
These states are nearly orthogonal (up to factors that decay as $e^{-\abs{\alpha}^{2}}$) and are steady states of a combination of four-photon driving and dissipation, since they are coherent linear superpositions of $0,4,8...$ and $2,6,10...$ photons, respectively. As with the two-photon states considered previously they are protected against phase noise by construction, and continuous driving ensures that drifting outside of the manifold of $\ket{\pm \alpha}$ and $\ket{\pm i \alpha}$ is energetically suppressed. We note the experimental realizations discussed in this work construct these states through tuned pulse sequences and coupling to a qubit rather than a four-photon drive itself; in absence of such driving phase noise can cause the state to drift out of the logical state manifold, though the high base coherence of 3d cavities makes this effect relatively weak. Like the two photon states, they are not protected against photon losses, but their more complex structure allows for efficient photon loss correction, as shown in FIG.~\ref{catECfig}.

We first notice that the parity operator $\hat{P} = \exp i \pi a_P^\dagger a_P$ returns 1 for both $\ket{0_L}$ and $\ket{1_L}$, so it does not distinguish them. Now imagine a photon has been lost; up to overall normalization this sends logical states the error states:
\begin{eqnarray}\label{errstates}
\ket{0_L} &\to& \ket{0_E} = 2 e^{-\abs{\alpha}^2 / 2} \sum_{n = 0}^{\infty} \frac{\alpha^{4n+3}}{\sqrt{\of{4n+3}!} } \ket{4n+3}, \\
\ket{1_L} &\to& \ket{1_E} = 2 e^{-\abs{\alpha}^2 / 2} \sum_{n = 0}^{\infty} \frac{\alpha^{4n+1}}{\sqrt{\of{4n+1}!} } \ket{4n+1}. \nonumber
\end{eqnarray}
If we measure $\hat{P}$ and a photon loss has occurred, we obtain $-1$ for either state, detecting that an error has occurred without distinguishing the two states. Leaving aside the issue of how to repair this lost photon for the moment, if we simply switch bases to the $E$ states from the $L$ states we find that an arbitrary superposition of $L$ states maps to an identical superposition of $E$ states under a single photon loss, without any relative phase accumulation or information loss. Now, if two photon losses occur in rapid succession then the two logical states will be mixed (causing a logical error), but if time between $\hat{P}$ measurements is small compared to the inverse loss rate this is rare and the effective lifetime of information encoded in the cat will be quadratically increased. 

To implement and measure $\hat{P}$ \cite{sunpetrenko2014} we dispersively couple the system to a superconducting qubit $Q$, with an effective coupling $\chi a_P^\dagger a_P a_Q^\dagger a_Q$. Assuming the qubit begins in $\ket{0}$ we apply a Hadamard gate to transform it to $\of{\ket{0} + \ket{1} }/\sqrt{2}$ and then wait for a time $t = \hbar \pi / \chi$ before applying a second Hadamard operation. If the parity of the resonator is even there is no phase accumulation on the qubit so it returns to $\ket{0}$, but if the resonator's parity is odd the qubit will now be in state $\ket{1}$, and measuring it thus measures the parity of the resonator. Single photon losses thus can be reliably detected from this mechanism, though the protocol is vulnerable to phase and loss errors in the qubit (which generally has much shorter coherence times than a 3d cavity) during the gate sequence. Nonetheless, it was sufficient to demonstrate QEC beyond break-even \cite{ofekpetrenko2016}. Future implementations of cat codes may be improved through more complex parity measurement schemes such as the one proposed by Cohen \textit{et al} \cite{cohensmith2016}. In that work, $\hat{P} = \chi \ket{1_Q} \bra{1_Q} \cos \pi a_P^\dagger a_P$ is implemented \textit{continuously}, rather than being the net result of a sequence of different operations, by quantum Zeno dynamics induced by two-photon driving and dissipation. This protocol does not entangle the transmon and the cavity so photon losses in the qubit do not disturb the cavity state, though a photon loss or phase error in the qubit during the operation can still cause an erroneous measurement result.

Depending on the implementation, various actions can be taken in response to the detection of a photon loss through a parity measurement. Experiments reported in the literature simply use the parity sequence as an error detection code, which still extends the lifetime of the encoded information as a detected photon loss has a deterministic effect on the logical manifold (\ref{catstates}) that preserves the relative amplitudes and phases of a superposition. One can do this whether the cat states are prepared by an initial pulse sequence and then left alone, or continuously refreshed by driving, and in this regime there are no beneficial dissipation sources integral to the logical qubit's operation. However, without four-photon driving \textit{and} four-photon dissipation the cat states will eventually decay to vacuum, an unrecoverable error even if each individual photon loss is successfully detected since all of the assumptions of orthogonality described above fail if $\alpha$ is too small. This precludes using cat qubits as part of a larger information processing system without incorporating driving and engineered dissipation to continuously ``re-inflate" the state. Likewise, depending on how gate protocols are to be engineered (see below) one may also need a mechanism for adding loss photons to return the system to the logical manifold after a single photon loss, as some gate protocols may not operate properly on odd parity states.

Logical operations on and between cat qubits, as well as schemes for extending the error correction cycle to recover from multiple photon losses and/or dephasing, are more complex subjects beyond the scope of this review. The cat code can be seen as a special case of a more general class of error correction codes for information encoded in harmonic oscillator states \cite{michaelsilveri2016}, which are capable of recovering from a larger range of error processes by entangling multiple resonators \cite{wanggao2016} or encoding more complex states in a single resonator (for example, two photon losses can be corrected if the states are encoded in superpositions of 6-photon states instead of 4-photon states). Logical operations can be applied through various methods \cite{mirrahimileghtas2014,heeresreinhold2016}, including adiabatic manipulations of the coherent state phases \cite{albertshu2016}. Gate sets which act as intended on the odd error manifold of cat states are particularly useful, since they can be enacted after an error is detected without having to replace the lost photon. However, designing a physical implementation of a complete, universal gate set, which loses no fidelity when a single photon is lost at any point during the gate, is a subtle challenge. Among other reasons, this is because any mechanisms which manipulate the cat state through interaction with a qubit are vulnerable to losses and dephasing in the qubit during the gate operation, making them no higher fidelity than ordinary qubit gates.

\begin{figure}
\includegraphics[width=3.25in]{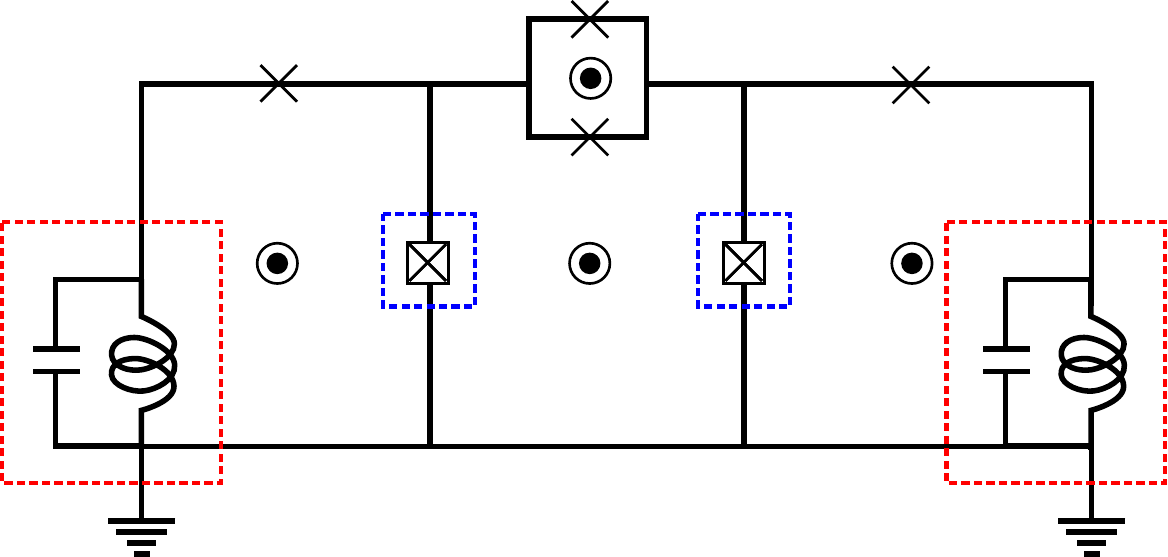}
\includegraphics[width=3.25in]{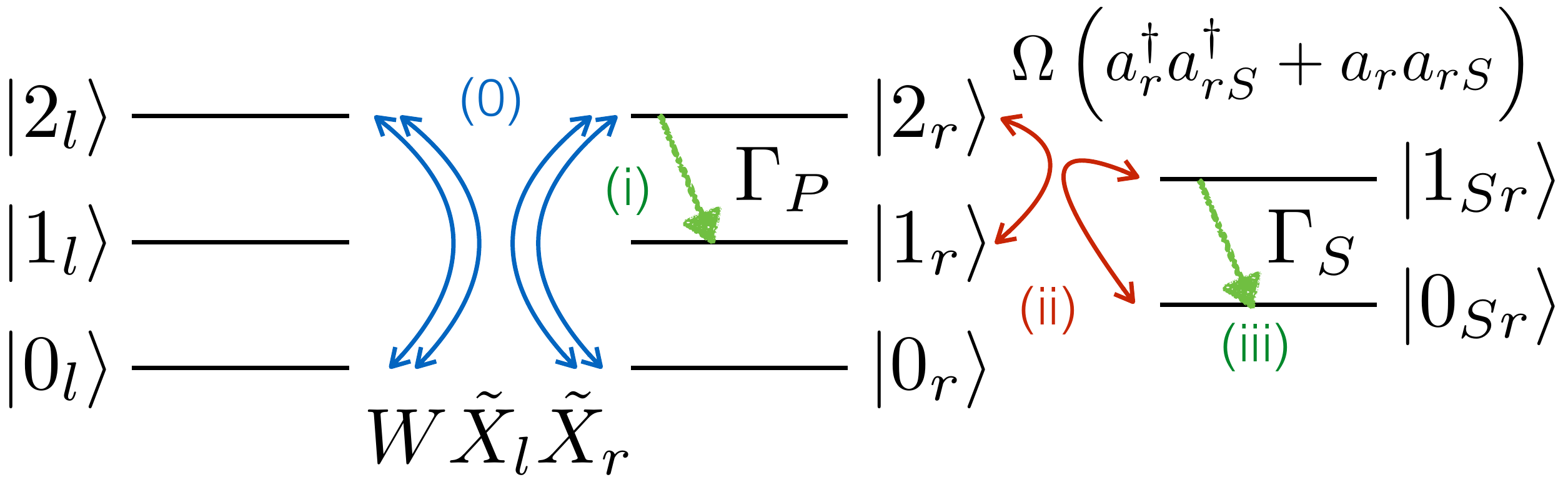}
\caption{Top: Circuit diagram for the Very Small Logical Qubit, taken from \cite{kapit2016}. The VSLQ consists of two transmon qubits, with bridged grounds, coupled by a flux driven junction or SQUID (essentially, a modified flux qubit). Each transmon is coupled to a lossy resonator through a high-frequency driven coupling element; these couplers drive a resonant two-photon process that corrects photon loss errors in the primary circuit. Bottom: error correction process in the VSLQ. The logical state manifold is spanned by superpositions of zero and two photon states in the left and right qubits with $\tX_l \tX_r \ket{L} =  \ket{L}$, stabilized by the parametric drive term $- W \tX_l \tX_r$ (blue arrows, step 0). The error correction ``cycle" begins at step (i) when a photon is lost from one of the qubits (rate $\Gamma_P$, green arrow). This state is resonant with an excitation in the corresponding lossy shadow qubit (or resonator), and the system Rabi-flops (step ii) under the two photon drive $\Omega \of{a_{r}^\dagger a_{rS}^\dagger + a_r a_{rS}}$ (red arrows), oscillating between a $\ket{1}$ state in the right qubit and the shadow qubit empty and returning the pair of qubits to the parent logical state $\ket{L}$ and exciting the shadow qubit (only the right shadow qubit is shown in the diagram). The shadow qubit photon rapidly decays (rate $\Gamma_S$, green arrow, step iii), returning the system to the pre-error state. At no point was the system measured by an external observer; the error correction is passive and automatic, though it cannot correct more than one photon loss at a time. The error correction drives and dissipation are always on in this system, and energetic resonance conditions ensure that the shadow qubits only impact the state of the primary qubit when a photon loss has occurred.}\label{VSLQcircuit}
\end{figure}

\subsection{Very small logical qubit}

The Very Small Logical Qubit (VSLQ) \cite{kapit2016}, itself inspired by the cat code, is an alternative proposal for building a logical qubit through engineered dissipation. While similar in basic mechanism-- both schemes rely on four-photon driving to create a degenerate manifold that allows for photon loss error correction-- their regimes of operation are very different. Where the cat code stores information in a high coherence resonator with a small nonlinearity, the VSLQ (which has yet to be realized experimentally) relies on strong driving in pairs of highly nonlinear qubits, with comparatively more rapid error correction and large drive amplitudes compensating for significantly lower base coherences.

The basic circuit for the VSLQ is shown in FIG.~\ref{VSLQcircuit}. The primary circuit consists of two transmon qubits with bridged grounds, coupled by a flux driven SQUID-- aside from the relative placement of the capacitances and values of the junctions it is identical to a four-junction flux qubit \cite{mooijorlando}. The SQUID is held at or near $\pi$ flux so that the two qubits are uncoupled in absence of oscillating tones. The two transmons $l$ and $r$ have different energies, and the flux driven coupler is driven at a pair of frequencies commensurate (the nonlinear junction coupling allows for frequency conversion and wave mixing) \cite{kapit2015} with two-photon exchange ($2 \omega_l - 2 \omega_r$) and four-photon pumping ($2 \omega_l + 2 \omega_r - 2 \delta$, where $-\delta$ is the nonlinearity of the transmon). To discuss the rotating frame VSLQ Hamiltonian, it is useful to define the following operators. Let $P_{a}^{b}$ be the projector onto states where object $a$ contains $b$ photons, and let $\tX_j \equiv \of{ \ket{2_j}\bra{0_j} + \ket{0_j} \bra{2_j}}$ and $\tZ_j \equiv P_j^2 - P_j^0$ be the Pauli $X$ and $Z$ operators in the $\ket{0}$ and $\ket{2}$ basis. The VSLQ Hamiltonian in the three-level basis of each transmon \footnote{The increasing nonlinearity allows us to ignore the $\ket{3}$ and $\ket{4}$ states; mixing with these states will produce small perturbative corrections that can be cancelled by altering the drive signal, and for simplicity we ignore them in this review.} is given by
\begin{eqnarray}\label{HVSLQ}
H_P = - W \tX_l \tX_r + \frac{\delta}{2} \of{P_{l}^1 + P_{r}^1}.
\end{eqnarray}
$H_P$ has a two-fold degenerate ground state, spanned by the product states where both $\tX_l$ and $\tX_r$ equal 1 or both equal -1, with no occupation of single photon $\ket{1}$ states. As remarked earlier, when in a logical state the VSLQ is protected against phase noise by the $W$ term, which suppresses low-frequency noise by giving phase errors a finite energy cost. Logical operators in this basis are $X_L = \tX_{l}$, $Z_{L} = \tZ_l \tZ_r$ and $Y_L = i X_L Z_L$; these operators all commute with $H_P$ and anticommute with each other, and partial combinations of them can implement any logical rotation desired.

The error correction ``cycle" in the VSLQ, shown in FIG.~\ref{VSLQcircuit}, is continuous and operates through engineered dissipation. Specifically, each of the two primary qubits is coupled to a lossy qubit or resonator (a ``shadow" object) through a parametric two-photon coupling $H_{PS} = \Omega \sum_{i=l,r} \of{a_{i}^\dagger a_{iS}^\dagger + a_{i} a_{iS} } $, and the shadow objects are detuned so that they have an excitation energy $\omega_S = \delta/2 + W$ in the rotating frame. As in the bit flip code example, we assume that $\delta \gg W \gg \Omega \sim \Gamma_S \gg \Gamma_P$, where $\Gamma_P = T_{1}^{-1}$ is the photon loss rate in the two qubits. Consequently, if no photons are lost the action of $H_{PS}$ creates two excitations at a total cost $W + \delta$, which is energetically forbidden, so it does nothing when acting on the logical state manifold. However, when a photon is lost, the ``hole" excitation in the primary qubit (a $\ket{1}$ state) is resonant with exciting the shadow object (and thus refilling the hole), which then rapidly decays. Because of the continuously operated $W$ term repairing the photon loss returns the system to the logical state manifold, as repairing to a state where $\tX_l \tX_r = -1$ is energetically forbidden, and the information contained in the undisturbed state of the other primary qubit is sufficient for this coupling to reconstruct the parent logical state. As no phase or rate differences are present between logical states where $\tX_l = +1$ or $\tX_r = -1$, the VSLQ fulfills all the conditions for passive quantum error correction.

\begin{figure}
\includegraphics[width=3.25in]{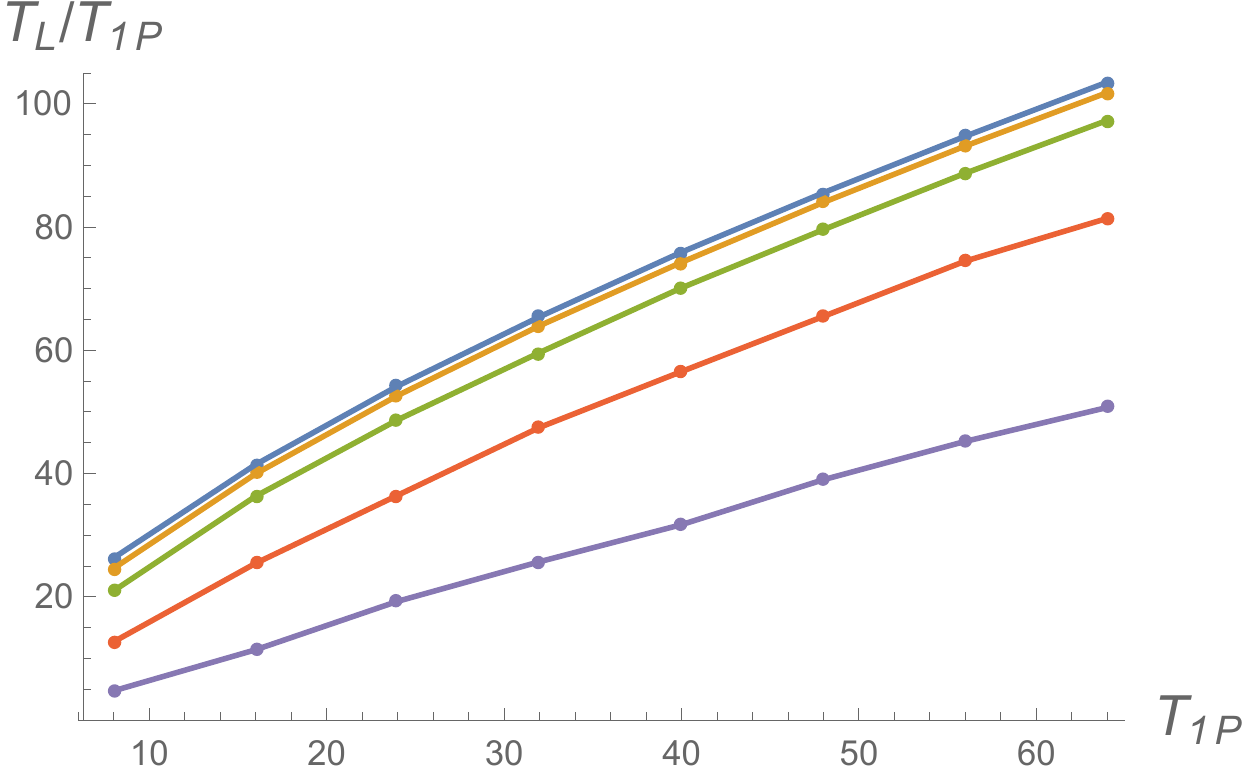}
\caption{Simulated lifetime of $\tX$ eigenstates in the VSLQ under photon losses and $1/f$ phase noise (E. Kapit, in perparation), for $W = 2 \pi \times 25$ MHz and $\delta = 2 \pi \times 300$ MHz, with $\Omega$ and $\Gamma_S$ chosen based on $T_{1P}$ to optimize lifetime. The plot demonstrates the extracted lifetime improvement, $T_L / T_{1P}$ of a $\tX$ eigenstate under photon losses with a rate $1/T_{1P}$. Each qubit also experiences $1/f$ phase with an average strength chosen such that the single qubit Ramsey $T_{2R}$ (free induction decay due to phase noise alone, assuming no photon loss) is infinite (blue) or $T_{2R} = \cuof{1,1/2,1/4,1/8} T_{1P}$ (top to bottom; gold, green, red and purple). Note that the transmons in the VSLQ experience twice the effective phase noise of a single qubit because the logical states are in the two-photon manifold. The lifetime is computed by numerically integrating the Lindblad equations with randomly fluctuating $h_{l/r} \of{t} a_{l/r}^\dagger a_{l/r}$ terms added, with the trajectory averaged over 400 random pairs of phase error signals per data point. The lifetime of $Y_L$ and $Z_L$ eigenstates is slightly less than half the $\tX$ lifetime as additional error channels contribute there. $\Omega$ and $\Gamma_S$ vary in the ranges $2\pi \times \cuof{2.63,1.24}$ and $\cuof{23.3,11.0}$, respectively; both decrease as $T_{1P}$ increases, reducing the contribution of errors induced by the shadow resonators themselves. }\label{1overf}
\end{figure}

Like the simplest versions of the cat code, the VSLQ cannot correct multiple simultaneous photon losses. If a second photon is lost from the same qubit before the first can be replaced, it produces a $\tX$ or $\tY$ operation with 50\% probability each; the former would dephase a superposition of $\tX$ eigenstates and the latter is an excursion from the logical state manifold that creates a logical error along any logical axis. Similarly, if a second photon is lost from the other qubit, the resulting $\ket{1_l 1_r}$ state will correct back to a random logical state and thus also causes a logical error. Nonetheless, if $\Omega \gg \Gamma_P$ these processes are comparatively rare and the lifetime can far exceed the base coherences $T_{1}$ and $T_{2}$. Simulated logical state lifetimes (against photon loss and $1/f$ phase noise) are shown in FIG.~\ref{1overf}.

As shown in the figure, it is comparatively easy to achieve very rapid error correction with the VSLQ, making it an attractive device proposal for near-term work on passive quantum error correction. That said, there are important challenges in ensuring its successful operation. Like the cat code, the VSLQ is unable to correct single photon \textit{addition} errors, though these events are comparatively rare. Further, the four-photon coupling must be carefully constructed to minimize static $\tZ_{l/r}$, $\tZ_{l} \tZ_{r}$ or $P_{l/r}^1 \tZ_{r/l}$ terms acting on the primary qubits. This is because when a photon is lost the $W$ term is effectively turned off and single qubit $\tZ_{l}$ or $\tZ_r$ terms can dephase the $\tX$ eigenstates, producing a logical error. Note that random weak $\tZ_{l/r}$ terms from $1/f$ noise decrease the lifetime through this channel, though not dramatically. Fortunately, the inherent tunability of the flux biased SQUID coupler allows these terms, which arise from stray capacitive interactions and off-resonant perturbative mixing with $\ket{3}$ and $\ket{4}$ states, to be perturbatively cancelled out by slightly adjusting the static flux bias or adding weak additional drive terms. Engineering a ``clean" continuous four-photon coupling would undoubtedly be the most challenging step in realizing the VSLQ in an experiment; all the other components have already been demonstrated in previous experiments discussed in this review. The VSLQ also supports a robust set of ``error-transparent" gate operations \cite{kapitetg2017}, which tolerate single photon losses that occur mid-gate. This allows for superlinear increases in one- and two-VSLQ gate fidelity with linear increases in base qubit $T_1$, 

\subsection{More complex models}

The cat code and VSLQ are extremely efficient for correcting single photon losses, but extending them to recover from two or more photon losses in rapid succession is difficult in the case of cat codes \cite{michaelsilveri2016} and likely impossible for the VSLQ \footnote{A three-qubit ring variant of the VSLQ could recover from two simultaneous losses in different qubits, but would still succumb to a logical error from two simultaneous losses in the same qubit.}. To correct larger sequences of errors it is natural to consider topological codes \cite{fowlersurface,terhal2015}. These codes simulate ``commuting stabilizer" Hamiltonians with degenerate ground states, where local errors cannot distinguish the states and instead create pairs of anyonic defects. To produce a logical error further local errors must separate the anyons and drag them across the system (producing a string operation), and if the local error rate is sufficiently slow, measurement based error correction can detect them and either track or correct the error processes themselves, protecting quantum information. This section considers passive, ``analog" versions of these digital codes.

An obvious-- if extraordinarily ambitious-- goal would be to construct a passive quantum error correction system capable of correcting arbitrarily long chains of errors if built from suitably many qubits, thus mitigating the need for any type of active error correction. While such a construction would be complex, the overhead requirements for active error correction through a surface code or similar scheme are themselves intimidating (for example, $10^6$ physical qubits with a cycle time of 1 $\rm{\mu s}$ produces a terabit of raw measurement data per second, which must be algorithmically decoded in real time). And as one would expect, designing a large scale quantum memory capable of full self-correction is an extremely difficult problem, for a number of important reasons that I will discuss only briefly here. Readers with a detailed interest in this subject should consider the extensive review article by Brown \textit{et al} \cite{brownloss2016}, which discusses the prospects and limitations of self-correcting quantum memory at finite temperature. 

The primary obstacle to engineering self-correcting quantum memory is physical: assuming the topological Hamiltonian can be described by a commuting stabilizer code with local interactions, no-go theorems \cite{nussinovortiz2008,nussinovortiz2009,nussinovortiz2009b,bravyiterhal2009,yoshida2011} rigorously rule out the existence of topological order at finite temperature in two dimensions. Self-correcting quantum memory is commonly thought to require an energy barrier between ground states that grows extensively with system size \cite{bacon2006,hammacastelnovo2009,chesirothlisberger2010,hutterwootton2012,pedrocchihutter2013,bombinchhajlany2013,wootton2013,beckertanamoto2013}. This is straightforward to achieve in classical memory: if you are reading this article from a computer hard drive, you are able to do so because simple 2d ferromagnets require extensive amounts of energy to flip mesoscopic domains, and domains are thus stable for exponentially long times below the transition temperature. However, for topological quantum memory, where states cannot be distinguished through sub-extensive sequences of operations, the largest energy barriers in 2d are O(1), though in 3d \textit{nearly} self-correcting memory is possible, as shown by \cite{bravyihaah2013}. Larger energy barriers can be introduced through long ranged interactions \cite{hammacastelnovo2009,chesirothlisberger2010,hutterwootton2012,pedrocchihutter2013,wootton2013,beckertanamoto2013,hutter2014}, though implementing these models requires complex perturbative gadget constructions \cite{kempekitaev,jordanfarhi2008,bravyidivincencenzo2008,ockoyoshida,beckertanamoto2013,bombinchhajlany2013,pedrocchihutter2013,hutter2014}, and are the resulting large energy barriers are generally not resilient to static defects \cite{landoncardinal2015}. A four dimensional variant of the toric code \cite{pastawskiclemente2011} is stable at finite temperature (and presumably could be protected against a white noise, ``infinite temperature" bath if the cooling drive were much more rapid), though four dimensional structures obviously present packaging challenges, to put it mildly.

In spite of these challenges, theoretical progress has been toward possible models for self-correcting quantum code in noisy environment. A host of long-ranged interacting models \cite{hammacastelnovo2009,chesirothlisberger2010,hutterwootton2012,pedrocchihutter2013,wootton2013,beckertanamoto2013,hutter2014} provide large energy barriers and could be implemented through complex gadget constructions. Fujii \textit{et al} \cite{fujiinegoro2014} devised a kind of hybrid digital-analog scheme where the output of a surface code is mapped to a 2d Ising ferromagnet and then cooled, mapping the resulting chains of operations back to the primary system to undo error chains. Kapit \textit{et al} proposed a fully passive surface code variant \cite{kapitchalker2015} using finite ranged interactions, which would be able to correct long but finite error chains of photon losses and/or phase errors and could be combined with active error correction. Though not directly related to self-correcting memory, Young and Sarovar proposed a passive error correction scheme compatible with adiabatic quantum computing \cite{youngsarovar2013} which is similar to some of these concepts. Further, the no-go theorems only apply to commuting stabilizer models, and more complex topological codes (where different parts of the Hamiltonian do not commute) could possibly display extensive energy barriers and quantum self correction; a subclass of these models called topological subsystem codes are reviewed in \cite{brownloss2016}. Lacking the commuting stabilizer structure these models can be much more difficult to study, however.

Though the above models are for the most part designed to correct errors occurring from a finite temperature bath, many of them could be adapted to protect against the white noise, effectively infinite temperature bath of photon losses as well. The reason for this is that with some exceptions, the extensive energy barriers arise from strong long range interactions that induce anyon confinement, so that separating a pair of anyons costs a finite amount of energy per step of separation. If the system is coupled to an engineered dissipation source targeted to remove energy in that range anyons will experience a continuous dissipative force that pushes them back together, which is able to correct long error chains so long as the energy cost per step does not vanish (in the work of Kapit \textit{et al} \cite{kapitchalker2015} this occurred when anyons were around 7 steps apart). This requirement-- that the energy cost per step must not vanish-- naturally implies an extensive energy barrier, and though satisfying it is very difficult, if it is satisfied the lifetime will increase exponentially with system size so long as the white noise error rate is slow compared to the dissipative repair rate.

\begin{figure}
\includegraphics[width=3.25in]{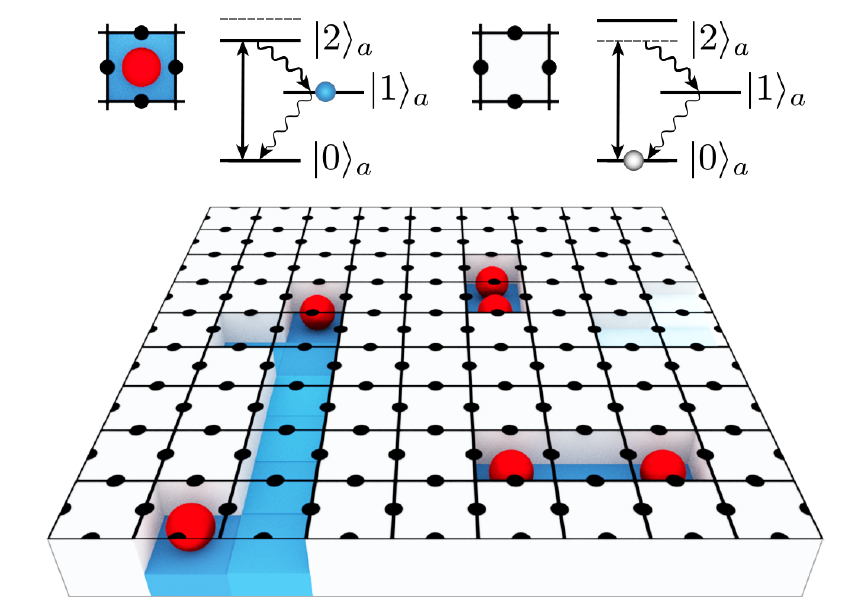}
\caption{Passive quantum error correction via engineered potential ``trenches," reproduced with permission from \cite{bardynkarzig2016}. In this scheme each plaquette of a surface code Hamiltonian \cite{kitaev,fowlersurface} is coupled to a  auxiliary system. As discussed in the text, a combination of engineered dissipation and strong interactions creates long-lived potential trenches that promote the recombination and annihilation of defects created by random interactions with the environment. Provided the errors themselves come from interactions with a cold bath, this can lead to exponential increases in the lifetime of quantum information encoded in the ground state.}\label{BKfig}
\end{figure}

Finally, a particularly intriguing proposal for achieving self-correction in a large scale system is by Bardyn and Karzig \cite{bardynkarzig2016}, who couple a surface code Hamiltonian \cite{kitaev,fowlersurface} to a set of three-level systems, as shown in FIG.~\ref{BKfig}. Random errors create plaquette or star defects, which then excite a three level system from the $\ket{0}$ to $\ket{2}$ state. Importantly, the resonant $\ket{0} \to \ket{2}$ driving only occurs where defects are present. As discussed earlier in the discussion of the subsequent work of Ma \textit{et al} \cite{maowens2017}, the $\ket{2} \to \ket{1}$ decay process is engineered to be rapid, but the $\ket{1}$ state is relatively long-lived, and when a local auxiliary system is in $\ket{1}$ it exerts a strong attractive potential on the topological defect. This in turn energetically suppresses further errors (a local error which would move a defect now costs finite energy; this process is ``free" in the toric code Hamiltonian itself) and thus impedes the anyon diffusion that causes logical errors. Further, defects which do diffuse tend to leave long ``trenches" of excited auxiliary states, within which motion is comparatively easy, promoting defect recombination. If anyonic defects share a spin they can be annihilated rapidly through coupling to a cold bath, leaving the isolated auxiliary $\ket{1}$ states to decay more slowly. Under the assumption that the base Hamiltonian is in the rest frame of the system (though the $\ket{0} \to \ket{2}$ process is driven) and errors come from interactions with a low-temperature bath, the resulting maximum lifetime scales exponentially in $1/T$ (though the system size must increase polynomially to achieve this), overcoming the ``no-go" theorems mentioned earlier through clever use of engineered dissipation. That said, a physical realization of the toric code Hamiltonian which does not rely on driving, or bring along with it strong 1- and 2-body correction terms (that could invalidate the results), is not obvious, but the necessary cold bath could be engineered in a driven system through broad-band engineered dissipation sources as discussed earlier \cite{kapitchalker2015,shabanineven2015}.

\section{Conclusions and Outlook}

The works of this review demonstrate the unique potential of engineered dissipation as a resource for quantum computing and simulation with superconducting devices. From single qubit stabilization to entanglement generation, many-body physics and quantum error corrections dissipative processes can generate complex dynamics completely autonomously, and often outperform measurement based schemes. As the field prepares to make the jump from few-qubit circuits to many-body systems and topological QEC, engineered dissipation could become an invaluable tool in realizing these lofty ambitions.

Going forward, I see two areas of research that are particularly promising. The first is in many-body physics; as argued earlier, dissipative schemes can stabilize gapped many-body ground states of strongly interacting Hamiltonians, even in cases where no digital protocol exists to generate, much less protect, the states. In comparison to more mature (in the many-body context) platforms such as cold atoms, superconducting qubits are extremely tunable at the individual qubit level and can implement nearly arbitrary pairwise interactions with the same physical hardware, though photon losses have thus far made it impossible to study true many-body states. Passive loss correction through colored bath engineering is an obvious solution to this problem. With photon loss tackled, superconducting circuits could probe many-body dynamics that are far beyond the scope of classical simulation.

Further, the dynamics of driven-dissipative systems are extremely complex, and it is obvious that driven-dissipative many-body systems (subject to white noise loss and colored baths for stabilization) can exhibit substantial entanglement. The large-$N$ behavior of these systems in two or more dimensions is hopelessly difficult to predict classically in the general case, but given that we expect non-classical correlations to persist against noise the resulting distributions may be a useful resource for approximate algorithms, such as machine learning.

Second, schemes such as the cat code and VSLQ clearly demonstrate that small circuits carefully tailored to the noise spectra of superconducting qubits can dramatically outperform stabilizer code error correction, at least at the single-error level. Similarly though the gadget constructions required are formidable in their own right, passive topological codes are capable of performing complex multi-error correction of long chains of quantum errors. Now, I find it extremely unlikely that a passive circuit will ever become efficient enough that \textit{no} digital error correction is required for a fault tolerant machine, but it is easy to imagine how engineered dissipation could be integrated into a measurement based code. For example, once suitable gate protocols are developed one can envision using cat states or VSLQs as drop-in replacements of transmons in a surface code, where the vast majority of local single device errors are passively corrected and only proportionally rarer multi-error events need to be tracked or eliminated through a stabilizer sequence. Of course, a great deal of work remains to design and rigorously benchmark such codes. To name just a few potential issues, one needs to tackle possible quantum correlations between errors, understand the impact of ``leakage" errors arising from the larger local Hilbert space, and devise gate protocols that minimize the impact of transient single-device errors on syndrome extraction. But if such efforts lead to a hybrid digital-analog code with reduced requirements in hardware and/or classical processing they would be well worth it and could speed the construction of a fault tolerant quantum computer.


\section{Acknowledgements}

I would like to thank Sergio Boixo, Eric Holland, Yao Lu, David Schuster, Jonathan Simon, and Vadim Smelyanskiy for useful discussions touching various points in this article. This work was supported by the Louisiana Board of Regents through grant (LEQSF(2016-19)-RD-A-19 and by the National Science Foundation through grant PHY-1653820.

\bibliography{EC_bib,SLbib}

\end{document}